%% file: main.tex
\documentclass[sigconf]{acmart}

\copyrightyear{2021}
\acmYear{2021}
\setcopyright{acmlicensed}
\acmConference[KDD '21]{Proceedings of the 27th ACM SIGKDD Conference on Knowledge Discovery and Data Mining}{August 14--18, 2021}{Virtual Event, Singapore}
\acmBooktitle{Proceedings of the 27th ACM SIGKDD Conference on Knowledge Discovery and Data Mining (KDD '21), August 14--18, 2021, Virtual Event, Singapore}
\acmPrice{15.00}
\acmDOI{10.1145/3447548.3467289}
\acmISBN{978-1-4503-8332-5/21/08}

\usepackage{algorithm}
\usepackage{algorithmic}
\usepackage{booktabs} 
\usepackage{amsfonts}
\usepackage{amsmath}
\usepackage{setspace}
\usepackage{multirow}
\usepackage{subfigure}
\usepackage{enumitem}
\usepackage{bm}
\usepackage{verbatim}

\begin{document}
\fancyhead{}

\newcommand{\ie}{\textit{i.e.}}


\title{Model-Agnostic Counterfactual Reasoning for Eliminating Popularity Bias in Recommender System}

\author{Tianxin Wei$^{1}$, Fuli Feng$^{2*}$, Jiawei Chen$^1$, Ziwei Wu$^1$, Jinfeng Yi$^3$ and Xiangnan He$^{1*}$}

\affiliation{\institution{$^1$University of Science and Technology of China, $^2$National University of Singapore, $^3$JD AI Research}}


\email{rouseau@mail.ustc.edu.cn, fulifeng93@gmail.com, cjwustc@ustc.edu.cn}
\email{maggiewuzw@gmail.com, yijinfeng@jd.com, xiangnanhe@gmail.com}

\thanks{* Fuli Feng and Xiangnan He are Corresponding Authors. }





\begin{spacing}{0.96}
\input{00_abstract.tex}
\input{00_CCSKeywords.tex}
\maketitle
\input{01_introduction.tex}
\input{03_preliminaries.tex}

\input{04_probemdefinition.tex}
\input{05_methodology.tex}

\input{06_experiments.tex}
\input{02_relatedwork.tex}
\input{07_conclusion.tex}


\begin{acks}
This work is supported by the National Natural Science Foundation of China (U19A2079, 61972372) and National Key Research and Development Program of China (2020AAA0106000).
\end{acks}
\end{spacing}

\bibliographystyle{ACM-Reference-Format}
\bibliography{ref}

\begin{spacing}{0.9}
\input{08_appendix}

\end{spacing}

\end{document}

%% file: 00_abstract.tex
\begin{abstract}

The general aim of the recommender system is to provide \textit{personalized} suggestions to users, which is opposed to suggesting \textit{popular} items. However, the normal training paradigm, \ie, fitting a recommender model to recover the user behavior data with pointwise or pairwise loss, makes the model biased towards popular items. This results in the terrible Matthew effect, making popular items be more frequently recommended and become even more popular. Existing work addresses this issue with Inverse Propensity Weighting (IPW), which decreases the impact of popular items on the training and increases the impact of long-tail items. Although theoretically sound, IPW methods are highly sensitive to the weighting strategy, which is notoriously difficult to tune.

In this work, we explore the popularity bias issue from a novel and fundamental perspective --- cause-effect. We identify that popularity bias lies in the \textit{direct effect} from the item node to the ranking score, such that an item's intrinsic property is the cause of mistakenly assigning it a higher ranking score. To eliminate popularity bias, it is essential to answer the counterfactual question that \textit{what the ranking score would be if the model only uses item property}. To this end, we formulate a causal graph to describe the important cause-effect relations in the recommendation process. During training, we perform multi-task learning to achieve the contribution of each cause; during testing, we perform counterfactual inference to remove the effect of item popularity. Remarkably, our solution amends the learning process of recommendation which is agnostic to a wide range of models --- it can be easily implemented in existing methods. We demonstrate it on Matrix Factorization (MF) and LightGCN~\cite{he2020lightgcn}, which are representative of the conventional and SOTA model for collaborative filtering. Experiments on five real-world datasets demonstrate the effectiveness of our method.


\begin{figure}[ttbp]
\setlength{\abovecaptionskip}{0.cm}
\includegraphics[width=0.8\linewidth]{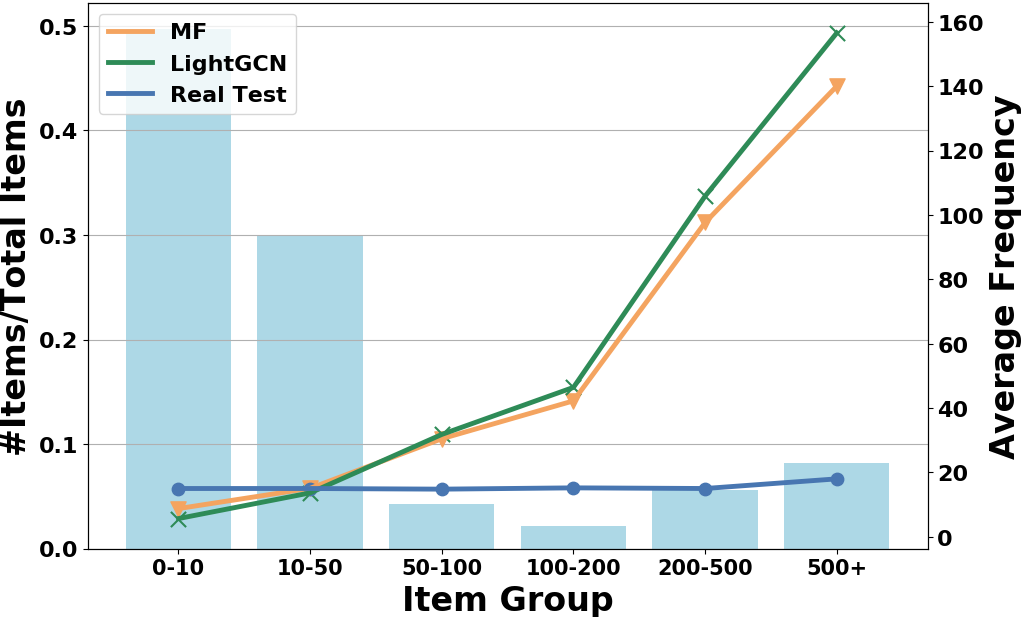}
\caption{An illustration of popularity bias in recommender system. Items are organized into groups w.r.t. the popularity in the training set wherein the background histograms indicate the ratio of items in each group, and the vertical axis represents the average recommendation frequency.}
\label{pop_bias}
\vspace{-0.5cm}
\end{figure}

\end{abstract}

%% file: 00_CCSKeywords.tex
\ccsdesc[500]{Information systems~Recommender systems}

\keywords{Recommendation, Popularity Bias, Causal Reasoning}

%% file: 01_introduction.tex
\section{Introduction}

Personalized recommendation has revolutionized a myriad of online applications such as e-commerce ~\cite{zhou2018deep,ying2018graph,wu2019neural}, search engines ~\cite{shen2005implicit}, and conversational systems ~\cite{sun2018conversational, lei2020estimation}. A huge number of recommender models~\cite{wang2019ngcf,kabbur2013fism,guo2015trustsvd} have been developed, for which the default optimization choice is reconstructing historical user-item interactions. However, the frequency distribution of items is never even in the interaction data, which is affected by many factors like exposure mechanism, word-of-mouth effect, sales campaign, item quality, etc. In most cases, the frequency distribution is long-tail, i.e., the majority of interactions are occupied by a small number of popular items. This makes the classical training paradigm biased towards recommending popular items, falling short to reveal the true preference of users~\cite{abdollahpouri2017controlling}.

Real-world recommender systems are often trained and updated continuously using real-time user interactions with training data and test data NOT independent and identically distributed (non-IID) \cite{zheng2020dice,algconfound}.
Figure \ref{pop_bias} provides an evidence of popularity bias on a real-world Adressa dataset~\cite{gulla2017adressa}, where we train a standard MF and LightGCN~\cite{he2020lightgcn} and count the frequency of items in the top-$K$ recommendation lists of all users. The blue line shows the item frequency of the real non-IID test dataset, which is what we expect.
As can be seen, more popular items in the training data are recommended much more frequently than expected, demonstrating a severe popularity bias. As a consequence, a model is prone to recommending items simply from their popularity, rather than user-item matching. 
This phenomenon is caused by the training paradigm, which identifies that recommending popular items more frequently can achieve lower loss thus updates parameters towards that direction. 
Unfortunately, such popularity bias will hinder the recommender from accurately understanding the user preference and decrease the diversity of recommendations. Worse still, the popularity bias 
will cause the Matthew Effect \cite{perc2014matthew} --- popular items are recommended more and become even more popular.

To address the issues of normal training paradigm, a line of studies push the recommender training to emphasize the long-tail items~\cite{liang2016modeling,bonner2018causal}. 
The idea is to downweigh the influence from popular items on recommender training, e.g., re-weighting their interactions in the training loss~\cite{liang2016causal,wang2018deconfounded}, incorporating balanced training data~\cite{bonner2018causal} or disentangling user and item embeddings~\cite{zheng2020dice}.
However, these methods lack fine-grained consideration of how item popularity affects each specific interaction, and a systematic view of the mechanism of popularity bias.
For instance, the interactions on popular items will always be downweighted than a long-tail item regardless of a popular item better matches the preference of the user.
We believe that instead of pushing the recommender to the long-tail in a blind manner, the key of eliminating popularity bias is to understand how item popularity affects each interaction.


\begin{figure}[ttbp]
\subfigure[User-item matching]{

\label{traditional}
\includegraphics[scale=0.15]{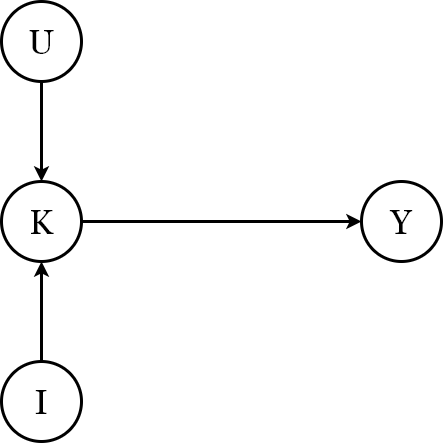}
}
\subfigure[Incorporating item popularity]{
\centering
\label{simple}
\includegraphics[scale=0.15]{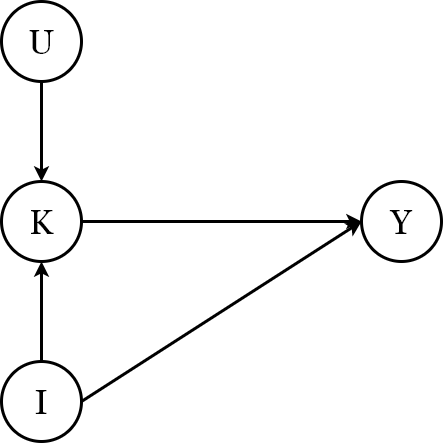}
}
\centering
\subfigure[Incorporating item popularity and user conformity]{
\centering
\label{true}
\includegraphics[scale=0.15]{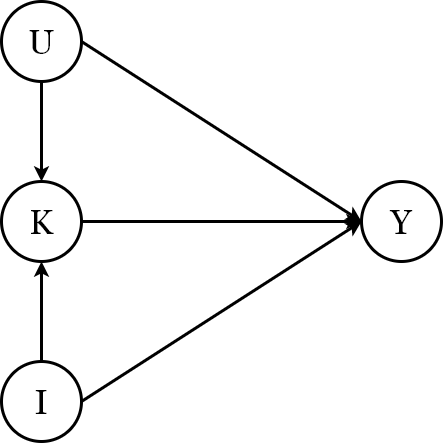}
}

\caption{Causal graph for (a) user-item matching; (b) incorporating item popularity; and (c) incorporating user conformity. I: item. U: user. K: matching features between user and item. Y: ranking score (e.g., the probability of interaction).}
\label{diff}
\vspace{-0.3cm}
\end{figure}
Towards this end, we explore the popularity bias from a fundamental perspective --- cause-effect, which has received little scrutiny in recommender systems. We first formulate a causal graph (Figure \ref{true}) to describe the important cause-effect relations in the recommendation process, which corresponds to the generation process of historical interactions.
In our view, three main factors affect the probability of an interaction: user-item matching, item popularity, and user conformity. However, existing recommender models largely focus on the user-item matching factor~\cite{he2017neural,DL4match} (Figure \ref{traditional}), ignoring how the item popularity affects the interaction probability (Figure \ref{simple}). Suppose two items have the same matching degree for a user, the item that has larger popularity is more likely to be known by the user and thus consumed. Furthermore, such impacts of item popularity could vary for different users, e.g., some users are more likely to explore popular items while some are not. As such, we further add a direct edge from the user node ($U$) to the ranking score ($Y$) to constitute the final causal graph (Figure \ref{true}). To eliminate popularity bias effectively, it is essential to infer the direct effect from the item node ($I$) to the ranking score ($Y$), so as to remove it during recommendation inference.

\begin{figure}[ttbp]
\includegraphics[width=0.98\linewidth]{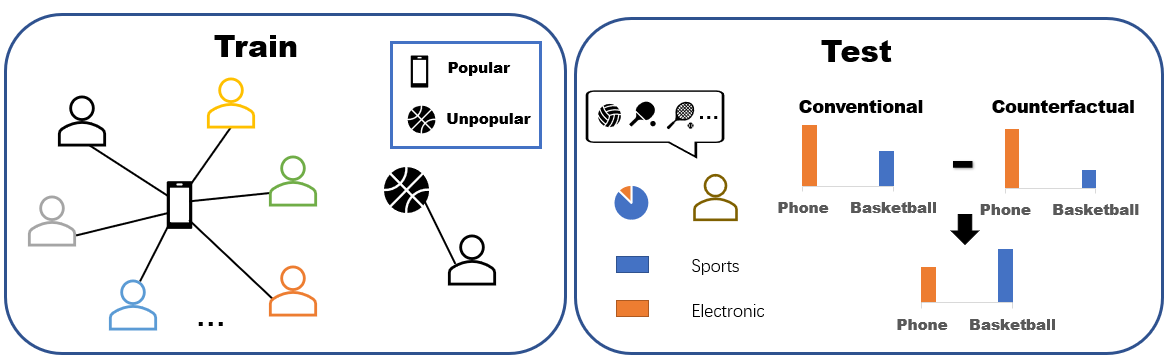}
\caption{An example of counterfactual inference.}
\label{example}
\vspace{-0.3cm}
\end{figure}

To this end, we resort to causal inference which is the science of analyzing the relationship between a cause and its effect~\cite{pearl2009causality}.
According to the theory of counterfactual inference~\cite{pearl2009causality}, the direct effect of $I\rightarrow Y$ can be estimated by imagining a world where the user-item matching is discarded, and an interaction is caused by item popularity and user conformity. 
To conduct popularity debiasing, we just deduct the ranking score in the counterfactual world from the overall ranking score.
Figure \ref{example} shows a toy example where the training data is biased towards iPhone, making the model score higher on iPhone even though the user is more interested in basketball.
Such bias is removed in the inference stage by deducting the counterfactual prediction.

In our method, to pursue a better learning of user-item matching, we construct two auxiliary tasks to capture the effects of $U \rightarrow Y$ and $I \rightarrow Y$. The model is trained jointly on the main task and two auxiliary tasks. 
Remarkably, our approach is model-agnostic and we implement it on MF \cite{koren2009matrix} and LightGCN \cite{he2020lightgcn} to demonstrate effectiveness.
To summarize, this work makes the following contributions:
\begin{itemize}[leftmargin=*]
\item Presenting a causal view of the popularity bias in recommender systems and formulating a causal graph for recommendation.
\item Proposing a model-agnostic counterfactual reasoning (MACR) framework that trains the recommender model according to the causal graph and performs counterfactual inference to eliminate popularity bias in the inference stage of recommendation.


\item Evaluating on five real-world recommendation datasets to demonstrate the effectiveness and rationality of MACR.
\end{itemize}

%% file: 03_preliminaries.tex

%% file: 04_probemdefinition.tex
\section{problem definition}
 Let $\mathcal{U} =\{u _1,u _2,...u _n\}$ and $\mathcal{I}  =\{i _1,i _2,...i _m\}$ denote the set of users and items, respectively, where $n$ is the number of users, and $m$ is the number of items.
The user-item interactions are represented by $\bm{Y} \in \mathbb{R}^{n\times m}$ where each entry,
\begin{equation}
    y_{ui}  = 
    \begin{cases}
    1,& \text{if user $u$ has interacted with item $i$,}\\
    0,& \text{otherwise.}
    \end{cases}
\end{equation}
The goal of recommender training is to learn a scoring function $f(u, i | \theta)$ from $Y$, which is capable of predicting the preference of a user $u$ over item $i$. Typically, the learned recommender model is evaluated on a set of holdout (e.g., randomly or split by time) interactions in the testing stage. However, the traditional evaluation may not reflect the ability to predict user true preference due to the existence of popularity bias in both training and testing.  
Aiming to focus more on user preference, we follow prior work~\cite{bonner2018causal,liang2016causal} to perform debiased evaluation where the testing interactions are sampled to be a uniform distribution over items.
This evaluation also can examine a model's ability in handling the popularity bias.


%% file: 05_methodology.tex
\section{methodology}
In this section, we first detail the key concepts about counterfactual inference (Section~3.1), followed by the causal view of the recommendation process (Section~3.2), the introduction of the MACR framework (Section~3.3), and its rationality for eliminating the popularity bias (Section~3.4). Lastly, we discuss the possible extension of MACR when the side information is available (Section~3.5).

\subsection{Preliminaries}\label{Preliminaries}

\begin{figure}[ttbp]
    \centering
    \includegraphics[width=1.0\linewidth]{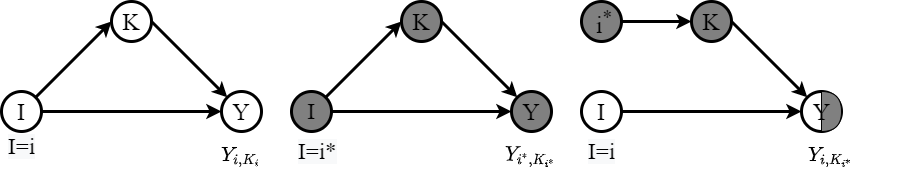}
    \vspace{-0.5cm}
    \caption{Example of causal graph where I, Y, and K denote cause, effect and mediator variable, respectively. Gray nodes mean the variables are at reference status (e.g., $I =i^*$).}
    \label{fig:preliminary}
    \vspace{-0.4cm}
\end{figure}

$\bullet$ \textit{Causal Graph.}
The \textit{causal graph} is a directed acyclic graph $G=\{V,E\}$, where $V$ denotes the set of variables and $E$ represents the cause-effect relations among variables \cite{pearl2009causality}. In a causal graph, a capital letter (e.g., $I$) denotes a variable and a lowercase letter (e.g., $i$) denotes its observed value. An edge means the ancestor node is a cause ($I$) and the successor node is an effect ($Y$).
Take Figure \ref{fig:preliminary} as an example, $I \rightarrow Y$ means there exists a direct effect from $I$ to $Y$. Furthermore, the path $I \rightarrow K \rightarrow Y$ means $I$ has an indirect effect on $Y$ via a mediator $K$.
According to the causal graph, the value of $Y$ can be calculated from the values of its ancestor nodes, which is formulated as:
\begin{equation}\label{eq:causal_cal}
Y_{i,k} = Y(I=i,K=k),
\end{equation}
where $Y(.)$ means the value function of $Y$. In the same way, the value of the mediator can be obtained through $k = K_i = K(I = i)$. In particular, we can instantiate $K(I)$ and $Y(I, K)$ as neural operators (e.g., fully-connected layers), and compose a solution that predicts the value of Y from input I.

$\bullet$ \textit{Causal Effect.}
The \textit{causal effect} of $I$ on $Y$ is the magnitude by which the target variable $Y$ is changed by a unit change in an ancestor variable $I$ \cite{pearl2009causality}. For example, the \textit{total effect} (TE) of $I = i$ on $Y$ is defined as:
\begin{equation}\label{eq:te}
    T E=Y_{i, K_{i}}-Y_{i^{*}, K_{i^{*}}} ,
\end{equation}
which can be understood as the difference between two hypothetical situations $I = i$ and $I = i^*$. $I=i^*$ refers to a the situation where the value of $I$ is muted from the reality, typically set the value as null. $K_{i^*}$ denotes the value of $K$ when $I=i^*$. Furthermore, according to the structure of the causal graph, TE can be decomposed into \textit{natural direct effect} (NDE) and \textit{total indirect effect} (TIE) which represent the effect through the direct path $I \rightarrow Y$ and the indirect path $I \rightarrow K \rightarrow Y$, respectively \cite{pearl2009causality}. 
NDE expresses the value change of $Y$ with $I$ changing from $i^*$ to $i$ on the direct path $I\rightarrow Y$ , while $K$ is set to the value when $I = i^*$, which is formulated as:
\begin{equation}
    N D E=Y_{i, K_{i^{*}}} - Y_{i^{*}, K_{i^{*}}},
\end{equation}
where $Y_{i, K_{i^{*}}} = Y(I=i,K=K(I=i^*))$. The calculation of $Y_i,K_{i^*}$ is a counterfactual inference since it requires the value of the same variable $I$ to be set with different values on different paths (see Figure~\ref{fig:preliminary}). Accordingly, 
TIE can be obtained by subtracting NDE from TE as following:
\begin{equation}
    T I E=T E - N D E=Y_{i, K_{i}}-Y_{i, K_{i^{*}}},
\end{equation}
which represents the effect of $I$ on $Y$ through the indirect path $I\rightarrow K\rightarrow Y$.

\subsection{Causal Look at Recommendation}
\label{sec:traditional}
In Figure \ref{traditional}, we first abstract the causal graph of most existing recommender models, where $U$, $I$, $K$, and $Y$ represent user embedding, item embedding, user-item matching features, and ranking score, respectively. The models have two main components: a matching function $K(U, I)$ that learns the matching features between user and item; and the scoring function $Y(K)$. For instance, the most popular MF model implements these functions as an element-wise product between user and item embeddings, and a summation across embedding dimensions. As to its neural extension NCF~\cite{he2017neural}, the scoring function is replaced with a fully-connected layer. Along this line, a surge of attention has been paid to the design of these functions. For instance, LightGCN~\cite{he2020lightgcn} and NGCF~\cite{wang2019ngcf} employ graph convolution to perform matching feature learning, ONCF~\cite{he2018outer} adopts convolutional layers as the scoring function. However, these models discards the user conformity and item popularity that directly affect the ranking score.

A more complete causal graph for recommendation is depicted in Figure~\ref{true} where the paths $U \rightarrow Y$ and $I \rightarrow Y$ represent the direct effects from user and item on the ranking score. A few recommender models follow this causal graph, e.g., the MF with additional terms of user and item biases~\cite{koren2009matrix} and NeuMF~\cite{he2017neural} which takes the user and item embeddings as additional inputs of its scoring function. While all these models perform inference with a forward propagation, the causal view of the inference over Figure \ref{traditional} and Figure \ref{true} are different, which are $Y_{K_{u, i}}$ and $Y_{u, i, K_{u, i}}$, respectively. However, the existing work treats them equally in both training and testing stages. For briefness, we use $\hat{y}_{ui}$ to represent the ranking score, which is supervised to recover the historical interactions by a recommendation loss such as the BCE loss~\cite{xue2017deep}:
\begin{equation}\label{eq:lo}
    L_{O}=\sum_{(u, i) \in D} -y_{ui}\log(\sigma(\hat{y}_{ui}))-(1-y_{ui})\log(1-\sigma(\hat{y}_{ui})), 
\end{equation}
where $D$ denotes the training set and $\sigma(\cdot)$ denotes the sigmoid function. $\hat{y}_{u,i}$ means either $Y_{K_{u,i}}$ or $Y_{u,i,K_{u,i}}$. In the testing stage, items with higher ranking scores are recommended to users.

Most of these recommender model suffer from popularity bias (see Figure \ref{pop_bias}). 
This is because $\hat{y}_{ui}$ is the likelihood of the interaction between user $u$ and item $i$, which is estimated from the training data and inevitably biased towards popular items in the data. From the causal perspective, item popularity directly affects $\hat{y}_{ui}$ via $I \rightarrow Y$, which bubbles the ranking scores of popular items. As such, blocking the direct effect from item popularity on $Y$ will eliminate the popularity bias.

\begin{figure}
\centering
\includegraphics[width=0.98\linewidth]{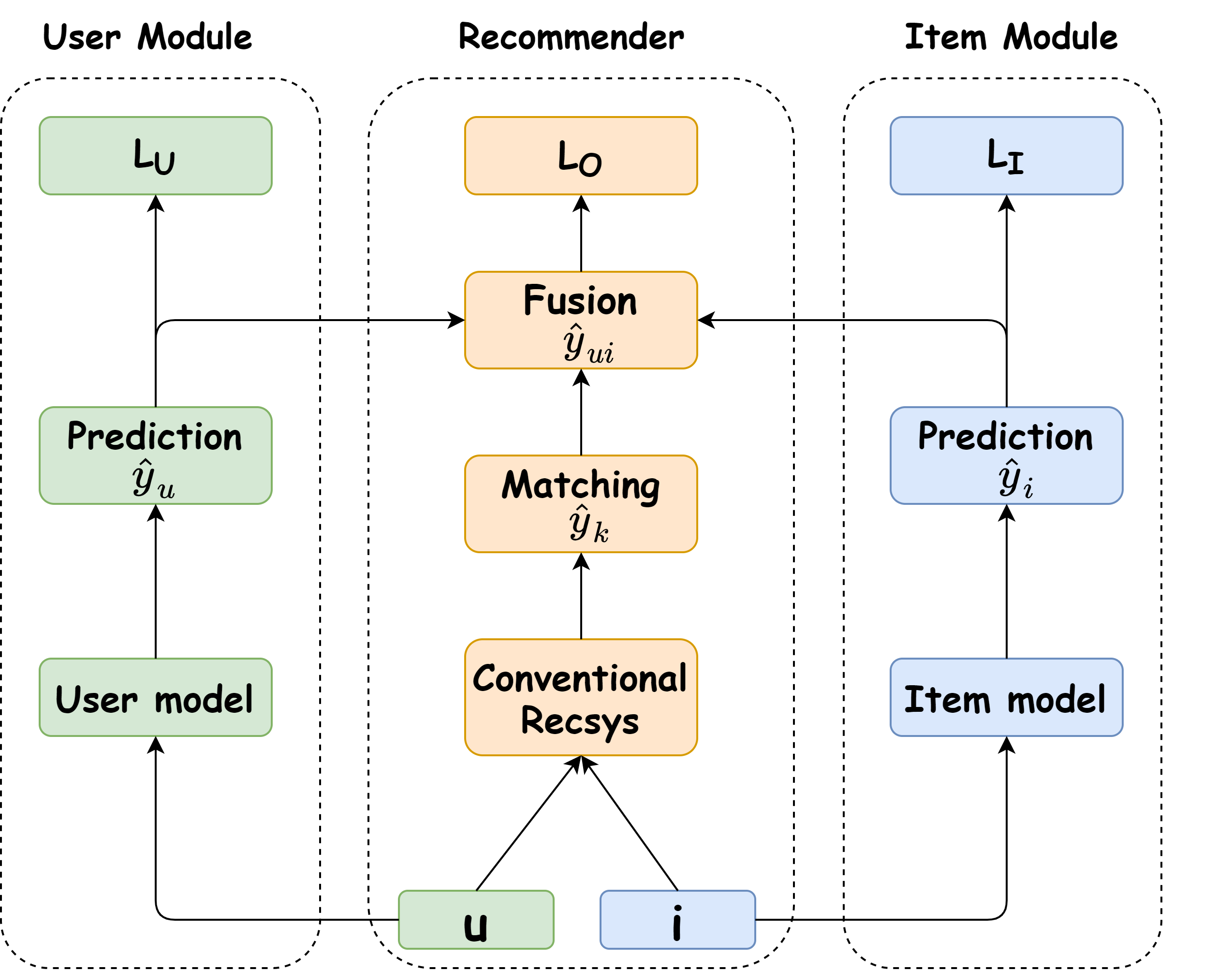}
\caption{The framework of MACR. The orange rectangles denote the main branch, i.e., the conventional recommender system. The blue and green rectangles denote the user and item modules, respectively.}
\label{framework}
\vspace{-0.725cm}
\end{figure}

\subsection{Model-Agnostic Counterfactual Reasoning}
To this end, we devise a model-agnostic counterfactual reasoning (MACR) framework, which performs multi-task learning for recommender training and counterfactual inference for making debiased recommendation. As shown in Figure~\ref{framework}, the framework follows the causal graph in Figure~\ref{true}, where the three branches correspond to the paths $U \rightarrow Y$, $U\&I \rightarrow K \rightarrow Y$, and $I \rightarrow Y$, respectively. This framework can be implemented over any existing recommender models that follow the structure of $U\&I \rightarrow K \rightarrow Y$ by simply adding a user module $Y_u(U)$ and an item module $Y_i(I)$. These modules project the user and item embeddings into ranking scores and can be implemented as multi-layer perceptrons. Formally, 
\begin{itemize}[leftmargin=*]
    \item \textit{User-item matching:} $\hat{y}_{k} = Y_k(K(U=u, I = i))$ is the ranking score from the existing recommender, which reflects to what extent the item $i$ matches the preference of user $u$.
    \item \textit{Item module:} $\hat{y}_{i} = Y_i(I = i)$ indicates the influence from item popularity where more popular item would have higher score.
    \item \textit{User module:} $\hat{y}_{u} = Y_u(U = u)$ shows to what extent the user $u$ would interact with items no matter whether the preference is matched. Considering the situation where two users are randomly recommended the same number of videos, one user may click more videos due to a broader preference or stronger conformity. Such ``easy'' user is expected to obtain a higher value of $\hat{y}_{u}$ and can be affected more by item popularity.
\end{itemize}
As the training objective is to recover the historical interactions $y_{ui}$, the three branches are aggregated into a final prediction score:
\begin{equation}\label{eq:ranking_score}
    \hat{y}_{ui} = \hat{y}_{k}*\sigma(\hat{y}_{i})*\sigma(\hat{y}_{u}),
\end{equation}
where $\sigma(\cdot)$ denotes the sigmoid function. It scales $\hat{y}_{u}$ and $\hat{y}_{i}$ to be click probabilities in the range of $[0, 1]$ so as to adjust the extent of relying upon user-item matching (\ie ~$\hat{y}_{k}$) to recover the historical interactions. For instance, to recover the interaction between an inactive user and unpopular item, the model will be pushed to highlight the user-item matching, \ie ~enlarging $\hat{y}_{k}$.

\paragraph{Recommender Training.} Similar to~\eqref{eq:lo}, we can still apply a recommendation loss over the overall ranking score $\hat{y}_{ui}$. To achieve the effect of the user and item modules, we devise a multi-task learning schema that applies additional supervision over $\hat{y}_{u}$ and $\hat{y}_{i}$. Formally, the training loss is given as:
\begin{equation}\label{ab}
    L=L_{O}+\alpha*L_I+\beta*L_U,
\end{equation}
where $\alpha$ and $\beta$ are trade-off hyper-parameters. Similar as $L_{O}$, $L_I$ and $L_U$ are also recommendation losses:
\begin{align}
    L_{U} &=\sum_{(u, i) \in D} -y_{ui}\log(\sigma(\hat{y}_{u}))-(1-y_{ui})\log(1-\sigma(\hat{y}_{u})), \notag \\
    L_{I} &=\sum_{(u, i) \in D} -y_{ui}\log(\sigma(\hat{y}_{i}))-(1-y_{ui})\log(1-\sigma(\hat{y}_{i})). \notag
\end{align}

\paragraph{Counterfactual Inference.} As aforementioned, the key to eliminate the popularity bias is to remove the direct effect via path $I \rightarrow Y$ from the ranking score $\hat{y}_{ui}$. To this end, we perform recommendation according to:
\begin{equation}\label{eq:ranking_test}
    \hat{y}_{k}*\sigma(\hat{y}_{i})*\sigma(\hat{y}_{u}) - c*\sigma(\hat{y}_{i})*\sigma(\hat{y}_{u}),
\end{equation}
where $c$ is a hyper-parameter that represents the reference status of $\hat{y}_{k}$. The rationality of the inference will be detailed in the following section. Intuitively, the inference can be understood as an adjustment of the ranking according to $\hat{y}_{ui}$. 
Assuming two items $i$ and $j$ with $\hat{y}_{ui}$ slightly lower than $\hat{y}_{uj}$, item $j$ will be ranked in front of $i$ in the common inference. Our adjustment will affect if item $j$ is much popular than $i$ where $\hat{y}_j >> \hat{y}_i$. Due to the subtraction of the second part, the less popular item $i$ will be ranked in front of j.. The scale of such adjustment is user-specific and controlled by $\hat{y}_u$ where a larger adjustment will be conducted for ``easy'' users.




\subsection{Rationality of the Debiased Inference}\label{cause-effect}
\begin{figure}[tbp]
\centering
\subfigure[Real world]{

\centering
\label{normal full}
\includegraphics[scale=0.2]{figures/causal.png}
}
\subfigure[Conterfactual world]{
\centering
\label{counterfactual full}
\includegraphics[scale=0.2]{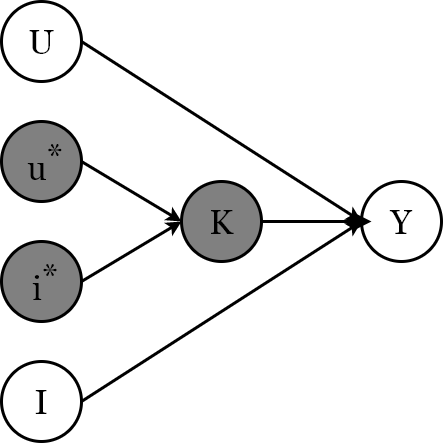}
}
\caption{Comparison between real world and counterfactual world causal graphs in recommender systems.}
\label{compare_full}
\vspace{-0.5cm}
\end{figure}

As shown in Figure \ref{true}, $I$ influences $Y$ through two paths, the indirect path $I \rightarrow K \rightarrow Y$ and the direct path $I \rightarrow Y$. 
Following the counterfactual notation in Section \ref{Preliminaries}, we 
calculate the NDE from $I$ to $Y$ through counterfactual inference where a counterfactual recommender system (Figure \ref{counterfactual full}) assigns the ranking score without consideration of user-item matching. As can be seen, the indirect path is blocked by feeding feature matching function $K(U, I)$ with the reference value of $I$, $K_{u^*, i^*}$. Formally, the NDE is given as:  
\begin{equation}
    NDE = Y(U=u,I=i, K=K_{u^*, i^*})-Y(U=u^*,I=i^*, K=K_{u^*, i^*}),\notag
\end{equation}
where $u^*$ and $i^*$ denote the reference values of $U$ and $I$, which are typically set as the mean of the corresponding variables, \ie ~the mean of user and item embeddings. 

According to Equation~\ref{eq:te}, the TE from $I$ to $Y$ can be written as:
\begin{equation}
    TE = Y(U=u,I=i, K=K_{u, i}) - Y(U=u^*,I=i^*, K=K_{u^*, i^*}).\notag
\end{equation}

Accordingly, eliminating popularity bias can be realized by reducing $NDE$ from $TE$, which is formulated as:
\begin{equation}\label{eq:causalrec}
    TE - NDE = Y(U=u,I=i, K=K_{u, i}) - Y(U=u,I=i, K=K_{u^*, i^*}),
\end{equation}
Recall that the ranking score is calculated according to Equation~\ref{eq:ranking_score}. As such, we have $Y(U=u,I=i, K=K_{u, i}) = \hat{y}_{k}*\sigma(\hat{y}_{i})*\sigma(\hat{y}_{u})$ and $Y(U=u,I=i, K=K_{u^*, i^*}) = c*\sigma(\hat{y}_{i})*\sigma(\hat{y}_{u})$ where $c$ denotes the value $\hat{y}_{k}$ with $K = K_{u^*, i^*}$. In this way, we obtain the ranking schema for the testing stage as Equation~\ref{eq:ranking_test}. Recall that $TIE = TE - NDE$, the key difference of the proposed counterfactual inference and normal inference is using TIE to rank items rather than TE. Algorithm in Appendix~\ref{procudure} describes the procedure of our method.


\subsection{Discussion}
There are usually multiple causes for one item click, such as items' popularity, category, and quality.
In this work, we focus on the bias revealed by the interaction frequency.
As an initial attempt to solve the problem from the perspective of cause-effect, we ignoring the effect of other factors. Due to the unavailability of side information~\cite{rendle2010factorization} on such factors or the exposure mechanism to uncover different causes for the recommendation, it is also non-trivial to account for such factors.

As we can access such side information, we can simply extend the proposed MACR framework by incorporating such information into the causal graph as additional nodes.
Then we can reveal the reasons that cause specific recommendations and try to further eliminate the bias, which is left for future exploration.

%% file: 06_experiments.tex
\section{experiments}
In this section, we conduct experiments to evaluate the performance of our proposed MACR. Our experiments are intended to answer the following research questions:
\begin{itemize}[leftmargin=*]
    \item \textbf{RQ1: }Does MACR outperform existing debiasing methods?
    \item \textbf{RQ2: }How do different hyper-parameter settings (e.g. $\alpha,\beta,c$) affect the recommendation performance?
    \item \textbf{RQ3: }How do different components in our framework contribute to the performance?
    \item \textbf{RQ4: }How does MACR eliminate the popularity bias?

\end{itemize}

\subsection{Experiment Settings}
\paragraph{Datasets}
Five real-world benchmark datasets are used in our experiments: ML10M is the widely-used~\cite{rendle2019difficulty,berg2017graph,zheng2016neural} dataset from MovieLens with 10M movie ratings. While it is an explicit feedback dataset, we have intentionally chosen it to investigate the performance of learning from the implicit signal. To this end, we transformed it into implicit data, where each entry is marked as 0 or 1 indicating whether the user has rated the item; Adressa \cite{gulla2017adressa} and Globo \cite{gabriel2019contextual} are two popular datasets for news recommendation; Also, the datasets Gowalla and Yelp from LightGCN \cite{he2020lightgcn} are used for a fair comparison. All the datasets above are publicly available and vary in terms of domain, size, and sparsity. The statistics of these datasets are summarized in Table~\ref{stat}. 
\begin{table}[ttbp]
\centering
\caption{Statistics of five different datasets.}
\vspace{-0.3cm}
\begin{tabular}{|l|l|l|l|l|}
\hline
         & Users    & Items   & Interactions     & Sparsity      \\ \hline
Adressa & 13,485  & 744   & 116,321  & 0.011594 \\ \hline
Globo    & 158,323 & 12,005 & 2,520,171 & 0.001326 \\ \hline
ML10M   & 69,166   & 8,790  & 5,000,415   & 0.008225 \\ \hline
Yelp    & 31,668 & 38,048 & 1,561,406 & 0.001300 \\ \hline
Gowalla    & 29,858 &  40,981 & 1,027,370 & 0.000840 \\
\hline

\end{tabular}
\label{stat}
\vspace{-0.3cm}
\end{table}

\begin{table*}[ttbp]
\caption{The performance evaluation of the compared methods with $K=20$. Rec means Recall. The bold-face font denotes the winner in that column. Note that the improvement achieved by MACR is significant ($p$-value $<<$ 0.05).}
\vspace{-0.3cm}
\resizebox{\textwidth}{29mm}{
\begin{tabular}{|c|l|l|l|l|l|l|l|l|l|l|l|l|l|l|l|}
\hline
\multicolumn{1}{|l|}{\multirow{2}{*}{}} & \multicolumn{3}{c|}{Adressa}                                                              & \multicolumn{3}{c|}{Globo}                                                                 & \multicolumn{3}{c|}{ML10M}                                                         & \multicolumn{3}{c|}{Yelp2018}                                                              & \multicolumn{3}{c|}{Gowalla}                                                               \\ \cline{2-16} 
  & \multicolumn{1}{c|}{HR} & \multicolumn{1}{c|}{Rec} & \multicolumn{1}{c|}{NDCG} & \multicolumn{1}{c|}{HR} & \multicolumn{1}{c|}{Rec} & \multicolumn{1}{c|}{NDCG} & \multicolumn{1}{c|}{HR} & \multicolumn{1}{c|}{Rec} & \multicolumn{1}{c|}{NDCG} & \multicolumn{1}{c|}{HR} & \multicolumn{1}{c|}{Rec} & \multicolumn{1}{c|}{NDCG} & \multicolumn{1}{c|}{HR} & \multicolumn{1}{c|}{Rec} & \multicolumn{1}{c|}{NDCG} \\ \hline \hline
MF                                      & 0.111 & 0.085 & 0.034 & 0.020 & 0.003 & 0.002 & 0.058 & 0.009 & 0.008 & 0.071 & 0.006 & 0.009 & 0.174 & 0.046 & 0.032                      \\ \hline
ExpoMF                                  & 0.112                    & 0.090                        & 0.037                      & 0.022                    & 0.005                        & 0.003                      & 0.061                    & 0.009                        & 0.008                      & 0.071                    & 0.006                        & 0.009                      & 0.175                    & 0.048                        & 0.034                      \\ \hline
CausE\_MF                               & 0.112                    & 0.084                         & 0.037                       & 0.023                    & 0.005                        & 0.003                      & 0.054                    & 0.008                        & 0.007                      & 0.066                    & 0.005                        & 0.008                      & 0.166                    & 0.045                        & 0.032                      \\ \hline
BS\_MF                                & 0.113 & 0.090 & 0.038 & 0.021 & 0.005 & 0.003 & 0.060 & 0.009 & 0.008 & 0.071 & 0.006 & 0.010 & 0.175 & 0.046 & 0.033                      \\ \hline
Reg\_MF                                 & 0.093 & 0.066 & 0.033 & 0.019 & 0.003 & 0.002 & 0.051 & 0.009 & 0.007 & 0.064 & 0.005 & 0.008 & 0.161 & 0.044 & 0.030                      \\ \hline
IPW\_MF                                 & 0.128 & 0.096 & 0.039 & 0.021 & 0.004 & 0.003 & 0.041 & 0.006 & 0.005 & 0.072 & 0.006 & 0.010 & 0.174 & 0.048 & 0.033                      \\ \hline
DICE\_MF                                & 0.133 & 0.098 & 0.041 & 0.033 & 0.007 & 0.006 & 0.055 & 0.011 & 0.007 & 0.082 & 0.008 & 0.011 & 0.177 & 0.052 & 0.033                      \\ \hline
MACR\_MF                              & \textbf{0.140}           & \textbf{0.109}               & \textbf{0.050}             & \textbf{0.112}           & \textbf{0.046}               & \textbf{0.026}             & \textbf{0.140}           & \textbf{0.041}               & \textbf{0.024}             & \textbf{0.135}           & \textbf{0.026}               & \textbf{0.019}             & \textbf{0.252}           & \textbf{0.077}               & \textbf{0.050}             \\ \hline \hline 
LightGCN                                & 0.123 & 0.098 & 0.040 & 0.017 & 0.005 & 0.003 & 0.038 & 0.006 & 0.005 & 0.061 & 0.004 & 0.009 & 0.172 & 0.045 & 0.032                      \\ \hline
CausE\_LightGCN                         & 0.115 & 0.082 & 0.037 & 0.014 & 0.005 & 0.003 & 0.036 & 0.005 & 0.005 & 0.061 & 0.005 & 0.009 & 0.173 & 0.046 & 0.033                      \\ \hline
BS\_LightGCN                           & 0.139 & 0.109 & 0.047 & 0.023 & 0.005 & 0.004 & 0.038 & 0.006 & 0.005 & 0.061 & 0.005 & 0.009 & 0.178 & 0.048 & 0.035                      \\ \hline
Reg\_LightGCN                            & 0.127 & 0.098 & 0.039 & 0.016 & 0.005 & 0.003 & 0.035 & 0.005 & 0.005 & 0.058 & 0.004 & 0.008 & 0.165 & 0.045 & 0.030                      \\ \hline
IPW\_LightGCN                          & 0.139 & 0.107 & 0.047 & 0.018 & 0.005 & 0.003 & 0.037 & 0.006 & 0.005 & 0.071 & 0.005 & 0.009 & 0.174 & 0.045 & 0.032                      \\ \hline
DICE\_LightGCN                                & 0.141 & 0.111 & 0.046 & 0.046 & 0.012 & 0.008 & 0.062 & 0.014 & 0.009 & 0.093 & 0.012 & 0.013 & 0.185 & 0.054 & 0.036                      \\ \hline
MACR\_LightGCN                        & \textbf{0.158}           & \textbf{0.127}               & \textbf{0.052}             & \textbf{0.132}           & \textbf{0.059}               & \textbf{0.030}             & \textbf{0.155}           & \textbf{0.049}               & \textbf{0.029}             & \textbf{0.148}           & \textbf{0.031}               & \textbf{0.018}             & \textbf{0.254}           & \textbf{0.077}               & \textbf{0.051}             \\ \hline
\end{tabular}}
\label{result}
\end{table*}

\paragraph{Evaluation.}\label{evaluation}
Note that the conventional evaluation strategy on a set of holdout interactions does not reflect the ability to predict user's preference, as it still follows the long tail distribution \cite{zheng2020dice}. Consequently, the test model can still perform well even if it only considers popularity and ignores users' preference \cite{zheng2020dice}. Thus, the conventional evaluation strategy is not appropriate for testing whether the model suffers from popularity bias, and we need to evaluate on the debiased data. To this end, we follow previous works \cite{liang2016causal,bonner2018causal,zheng2020dice} to simulate debiased recommendation where the testing interactions are sampled to be a uniform distribution over items. In particular, we randomly sample 10\% interactions with equal probability in terms of items as the test set, another 10\% 
as the validation set, and leave the others as the biased training data\footnote{We refer to \cite{liang2016causal,bonner2018causal,zheng2020dice} for details on extracting an debiased test set from biased data.}. 
%
%
%
%
%
We report the all-ranking performance w.r.t. three widely used metrics: Hit Ratio (HR), Recall, and Normalized Discounted Cumulative Gain (NDCG) cut at $K$.


\subsubsection{Baselines}
We implement our MACR with the classic MF (MACR\_MF) and the state-of-the-art LightGCN (MACR\_LightGCN) to explore how MACR boosts recommendation performance. We compare our methods with the following baselines:
\begin{itemize}[leftmargin=*]
    \item \textbf{MF \cite{koren2009matrix}: } This is a representative collaborative filtering model as formulated in Section \ref{sec:traditional}.
    \item \textbf{LightGCN~\cite{he2020lightgcn}: } This is the state-of-the-art collaborative filtering recommendation model based on light graph convolution as illustrated in Section~\ref{sec:traditional}.
    \item \textbf{ExpoMF~\cite{liang2016modeling}: } A probabilistic model that separately estimates the user preferences and the exposure.
    \item \textbf{CausE\_MF, CausE\_LightGCN~\cite{bonner2018causal}: } CausE is a domain adaptation algorithm that learns from debiased datasets to benefit the biased training. In our experiments, we separate the training set into debiased and biased ones to implement this method. Further, we apply CausE into two recommendation models (i.e. MF and LightGCN) for fair comparisons. Similar treatments are used for the following debias strategy.
    \item \textbf{BS\_MF, BS\_LightGCN \cite{koren2009matrix}: } BS learns a biased score from the training stage and then remove the bias in the prediction in the testing stage. The prediction function is defined as:$\hat{y}_{ui}=\hat{y}_{k}+b_i$, where $b_i$ is the bias term of the item $i$.
    \item \textbf{Reg\_MF, Reg\_LightGCN \cite{abdollahpouri2017controlling}: } Reg is a regularization-based approach that intentionally downweights the short tail items, covers more items, and thus improves long tail recommendation.
    \item \textbf{IPW\_MF, IPW\_LightGCN: \cite{liang2016causal,schnabel2016recommendations} } IPW Adds the standard Inverse Propensity Weight to reweight samples to alleviate item popularity bias.
    \item \textbf{DICE\_MF, DICE\_LightGCN: \cite{zheng2020dice}} This is a state-of-the-art method for learning causal embedding to cope with popularity bias problem. It designs a framework with causal-specific data to disentangle interest and popularity into two sets of embedding. We used the code provided by its authors.

\end{itemize}
As we aim to model the interactions between users and items, we do not compare with models that use side information. We leave out the comparison with other collaborative filtering models, such as NeuMF \cite{he2017neural} and NGCF~\cite{wang2019ngcf}, because LightGCN \cite{he2020lightgcn} is the state-of-the-art collaborative filtering method at present. Implementation details and detailed parameter settings of the models can be found in Appendix~\ref{parametersetting}.


\subsection{Results (RQ1)}
\label{sec:result}
Table ~\ref{result} presents the recommendation performance of the compared methods in terms of HR@20, Recall@20, and NDCG@20. The boldface font denotes the winner in that column. Overall, our MACR consistently outperforms all compared methods on all datasets for all metrics. The main observations are as follows:

\begin{itemize}[leftmargin=*]
    \item In all cases, our MACR boosts MF or LightGCN by a large margin. Specifically, the average improvement of MACR\_MF over MF on the five datasets is 153.13\% in terms of HR@20 and the improvement of MACR\_LightGCN over LightGCN is 241.98\%, which are rather substantial. These impressive results demonstrate the effectiveness of our multi-task training schema and counterfactual reasoning, even if here we just use the simple item and user modules. MACR potentially can be further improved by designing more sophisticated models.
    
    
    \item In most cases, LightGCN performs worse than MF, but in regular dataset splits, as reported in \cite{he2020lightgcn}, LightGCN is usually a performing-better approach. As shown in Figure \ref{pop_bias}, with the same training set, we can see that the average recommendation frequency of popular items on LightGCN is visibly larger than MF. This result indicates that LightGCN is more vulnerable to popularity bias. The reason can be attributed to the embedding propagation operation in LightGCN, where the influence of popular items is spread on the user-item interaction graph which further amplifies the popularity bias. However, in our MACR framework, MACR\_LightGCN performs better than MACR\_MF. This indicates that our framework can substantially alleviate the popularity bias. 
    \item In terms of datasets, we can also find that the improvements over the Globo dataset are extremely large. This is because Globo is a large-scale news dataset, and the item popularity distribution is particularly skewed. Popular news in Globo is widely read, while some other unpopular news has almost no clicks. This result indicates our model's capability of addressing popularity bias, especially on long-tailed datasets.
    \item As to baselines for popularity debias, Reg method \cite{abdollahpouri2017controlling} have limited improvement over the basic models and even sometimes perform even worse. The reason is that Reg simply downweights popular items without considering their influence on each interaction. CausE also performs badly sometimes as it relies on the debiased training set, which is usually relatively small and the model is hard to learn useful information from. BS and IPW methods can alleviate the bias issue to a certain degree. DICE achieved the best results among the baselines. This indicates the significance of considering popularity as a cause of interaction.
\end{itemize}
In Appendix~\ref{metric}, we also report our experimental results on Adressa dataset w.r.t. different values of $K$ in the metrics for more comprehensive evaluation.

\begin{figure}[ttbp]
\subfigure[MACR\_LightGCN]{

\includegraphics[scale=0.2]{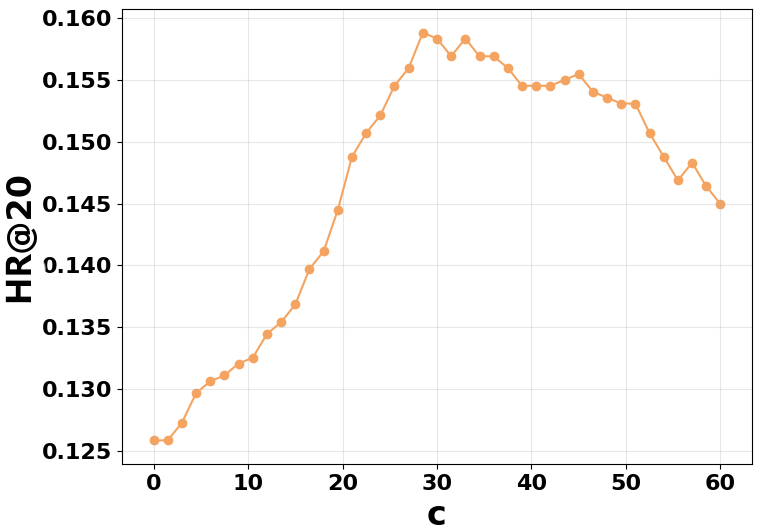}
}
\subfigure[MACR\_MF]{
\centering
\includegraphics[scale=0.2]{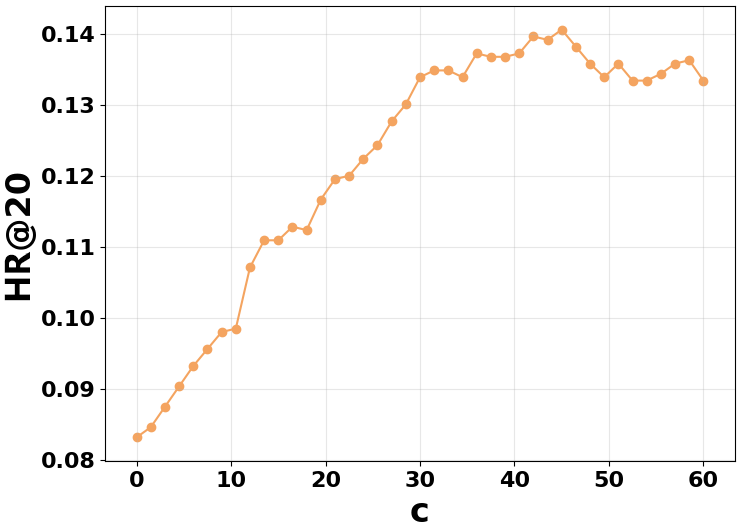}
}
\centering
\vspace{-0.5cm}
\caption{Effect of $c$ on MACR\_LightGCN and MACR\_MF w.r.t HR@20.}
\label{diffc}
\end{figure}

\subsection{Case Study}
\subsubsection{Effect of Hyper-parameters (RQ2)}
Our framework has three important hyper-parameters, $\alpha$, $\beta$, and $c$. Due to space limitation, we provide
the results of parameter sensitivity analysis of $\alpha$, $\beta$ in Appendix~\ref{hyper}.

The hyper-parameter $c$ as formulated in Eq.~\eqref{eq:ranking_test}
controls the degree to which the intermediate matching preference is blocked in prediction. We conduct experiments on the Adressa dataset on MACR\_LightGCN and MACR\_MF and test their performance in terms of HR@20. As shown in Figure~\ref{diffc}, taking MACR\_LightGCN as an instance, as $c$ varies from 0 to 29, 
the model performs increasingly better while further increasing $c$ 
is counterproductive. This illustrates that the proper degree of blocking intermediate matching preference benefits the popularity debias and improves the recommendation performance.

Compared with MACR\_MF, MACR\_LightGCN is more sensitive to $c$, as its performance drops more quickly after the optimum. It indicates that LightGCN is more vulnerable to popularity bias, which is consistent with our findings in Section~\ref{sec:result}.

\subsubsection{Effect of User Branch and Item Branch (RQ3)}
Note that our MACR not only incorporates user/item's effect in the loss function but also fuse them in the predictions. To investigate the integral effects of user and item branch, we conduct ablation studies on MACR\_MF on the Adressa dataset and remove different components at a time for comparisons. Specifically, we compare MACR with its four special cases: MACR\_MF w/o user (item) branch, where user (or item) branch has been removed; MACR\_MF w/o $L_I$ ($L_U$), where we just simply remove $L_I$ ($L_U$) to 
block the effect of user (or item) branch on training but retain their effect on prediction.

From Table~\ref{bracheff} we can find that both user branch and item branch boosts recommendation performance. Compared with removing the user branch, the model performs much worse when removing the item branch. Similarly, compared with removing $L_{U}$, removing $L_{I}$ also harms the performance more heavily. This result validates that item popularity bias has more influence than user conformity on the recommendation.

Moreover, compared with simply removing $L_{I}$ and $L_{U}$, removing the user/item branch makes the model perform much worse. This result validates the significance of further fusing the item and user influence in the prediction.

\begin{table}[ttbp]
\caption{Effect of user and item branch on MACR\_MF.}
\vspace{-0.3cm}
\resizebox{0.45\textwidth}{!}{
\begin{tabular}{|c|l|l|l|}
\hline
\multicolumn{1}{|l|}{}     & \multicolumn{1}{c|}{HR@20} & \multicolumn{1}{c|}{Recall@20} & \multicolumn{1}{c|}{NDCG@20} \\ \hline \hline
MACR\_MF                 & \textbf{0.140}           & \textbf{0.109}               & \textbf{0.050}             \\ \hline
MACR\_MF w/o user branch & 0.137                    & 0.106                        & 0.046                      \\ \hline
MACR\_MF w/o item branch & 0.116                    & 0.089                        & 0.038                      \\ \hline
MACR\_MF w/o $L_{I}$ & 0.124                    & 0.096                        & 0.043
\\ \hline
MACR\_MF w/o $L_{U}$ & 0.138                      & 0.108                 &0.048
\\ \hline
\end{tabular}}
\label{bracheff}
\end{table}

\begin{figure}[ttbp]
\subfigure[LightGCN]{

\includegraphics[scale=0.19]{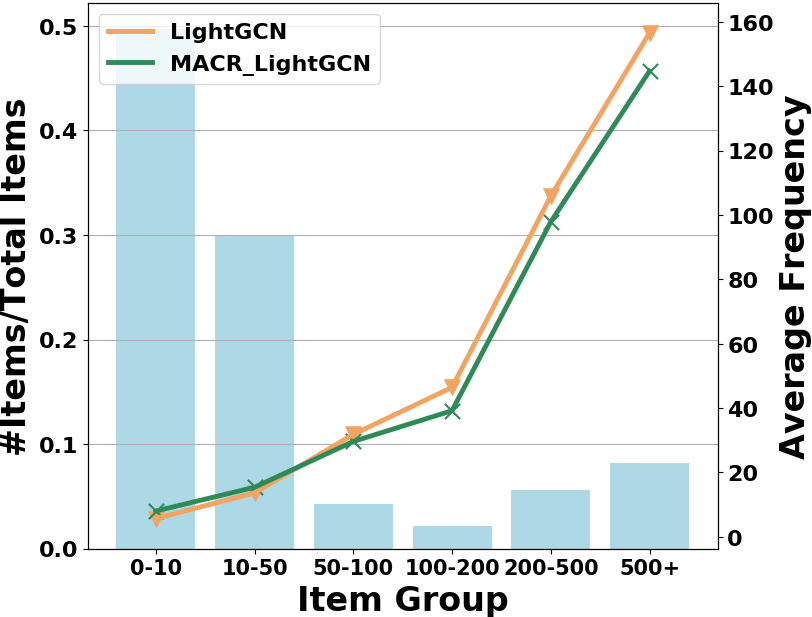}
}
\subfigure[MF]{
\centering
\includegraphics[scale=0.19]{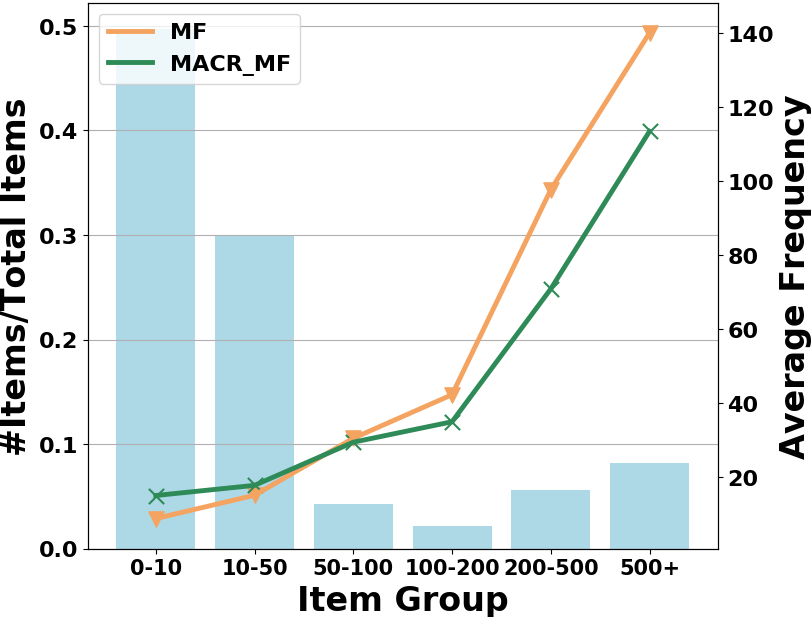}
}
\centering
\vspace{-0.5cm}
\caption{Frequency of different item groups recommended by LightGCN (MF) and MACR\_LightGCN (MACR\_MF).}
\vspace{-0.4cm}
\label{diff_causal}
\end{figure}

\begin{figure}[ttbp]
\subfigure[LightGCN]{

\includegraphics[scale=0.18]{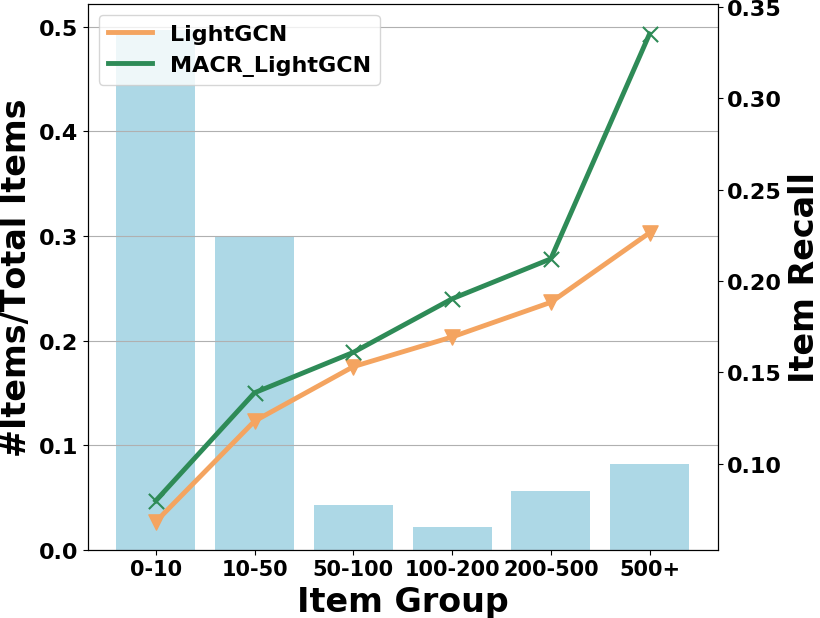}
}
\subfigure[MF]{
\centering
\includegraphics[scale=0.18]{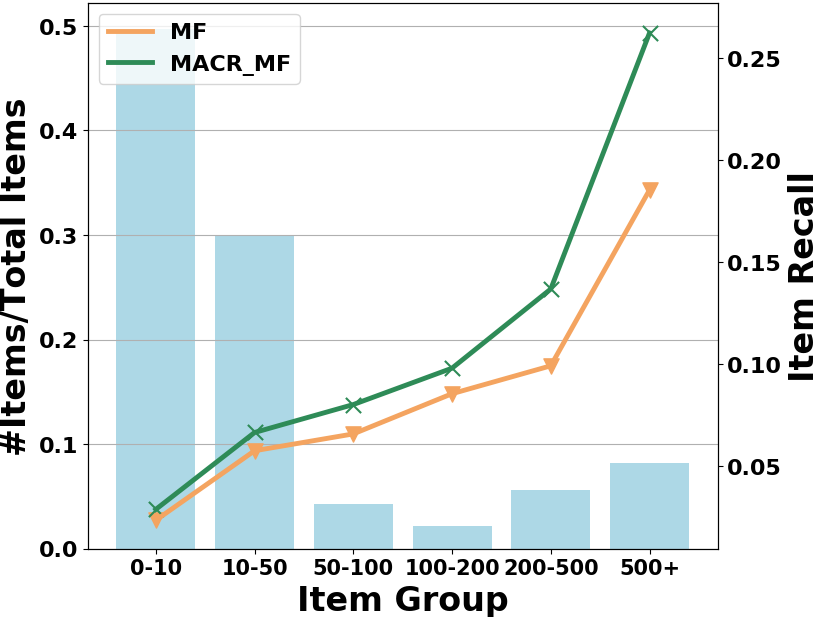}
}
\centering
\vspace{-0.5cm}
\caption{Average item recall in different item groups on Adressa.}
\label{item_acc}
\vspace{-0.6cm}
\end{figure}

\subsubsection{Debias Capability (RQ4)}
We then investigate whether our model alleviates the popularity bias issue. We compare MACR\_MF and MACR\_LightGCN with their basic models, MF and LightGCN. As shown in Figure~\ref{diff_causal}, we show the recommendation frequency of different item groups. We can see that our methods indeed reduce the recommendations frequency of popular items and recommend more items that are less popular. Then we conduct in Figure~\ref{item_acc} an experiment to show the item recommendation recall in different item groups. In this experiment, we recommend each user 20 items and calculate the item recall. If an item appears $N$ times in the test data, its item recall is the proportion of it being accurately recommended to test users. We have the following findings. 
\begin{itemize}[leftmargin=*]
    \item The most popular item group has the greatest recall increase, but our methods in Figure \ref{diff_causal} show the recommendations frequency of popular items is reduced. It means that traditional recommender systems (MF, LightGCN) are prone to recommend more popular items to unrelated users due to popularity bias. In contrast, our MACR reduces the item's direct effect and recommends popular items mainly to suitable users. This confirms the importance of matching users and items for personalized recommendations rather than relying on item related bias.
    \item The unpopular item group has relatively small improvement. This improvement is mainly due to the fact that we recommend more unpopular items to users as shown in Figure \ref{diff_causal}. Since these items rarely appear in the training set, it is difficult to obtain a comprehensive representation of these items, so it is difficult to gain a large improvement in our method.
\end{itemize}

\begin{figure}[ttbp]
\subfigure[LightGCN]{

\includegraphics[scale=0.18]{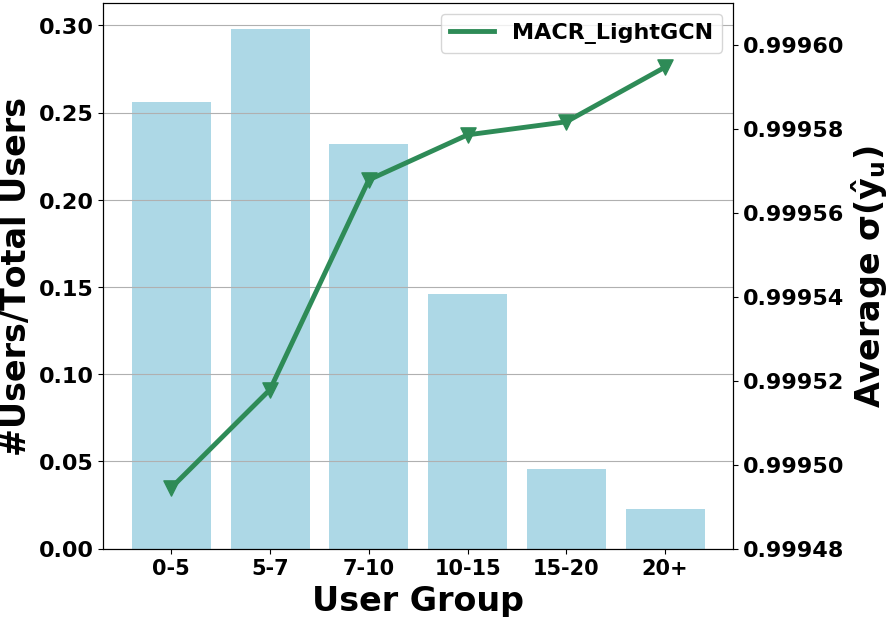}
}
\subfigure[MF]{
\centering
\includegraphics[scale=0.18]{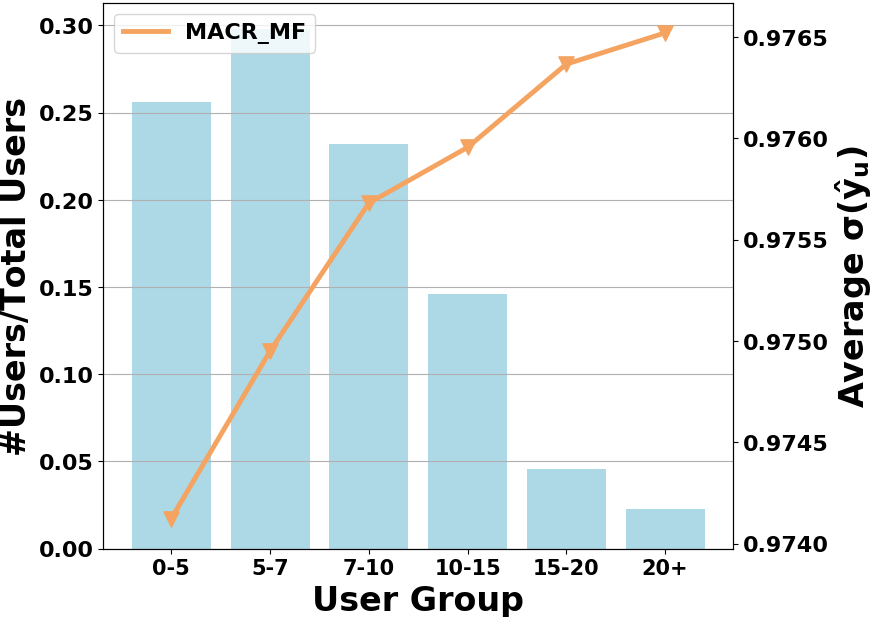}
}
\vspace{-0.6cm}
\centering
\caption{Average $\sigma(\hat{y}_u)$ comparison for different user groups on Adressa.}
\label{diffyu}
\vspace{-0.6cm}
\end{figure}

\begin{figure}[ttbp]
\subfigure[LightGCN]{

\includegraphics[scale=0.18]{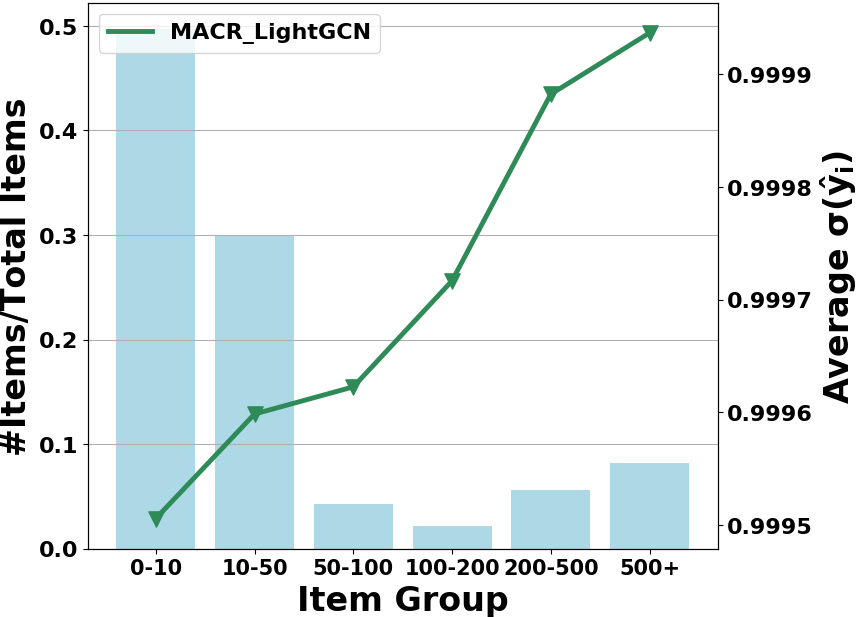}
}
\subfigure[MF]{
\centering
\includegraphics[scale=0.18]{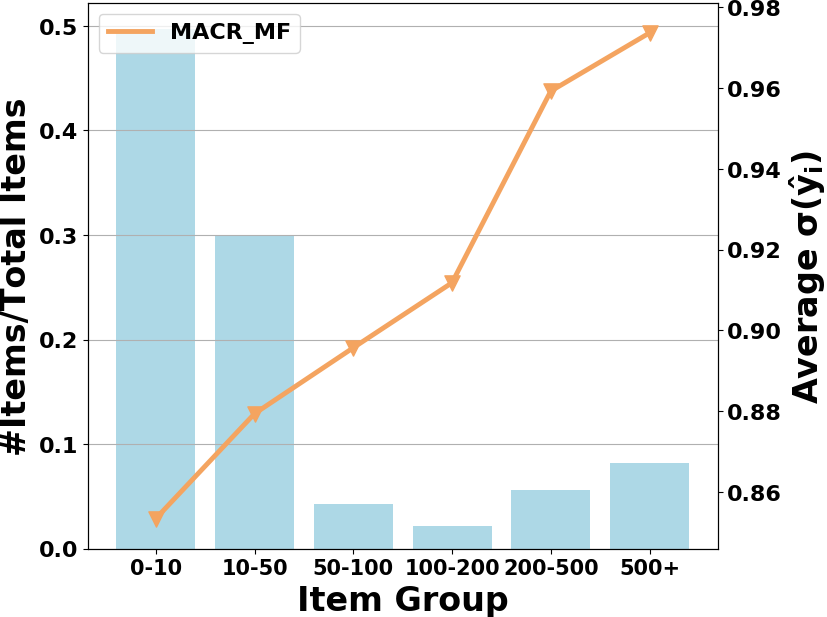}
}
\centering
\vspace{-0.5cm}
\caption{Average $\sigma(\hat{y}_i)$ comparison for different item groups on Adressa.}
\label{diffyi}
\end{figure}

To investigate why our framework benefits the debias in the recommendation, we explore what user branch and item branch, i.e., $\hat{y}_u$ and $\hat{y}_i$, actually learn in the model.
We compare $\sigma(\hat{y}_u)$ and $\sigma(\hat{y}_i)$ as formulated in Eq.~\eqref{eq:ranking_score} , which is the output for the specific user $u$ or item $i$ from the user/item model after the sigmoid function, capturing user conformity and item popularity in the dataset. 
In Figure~\ref{diffyu}, the background histograms indicate the proportion of users in each group involved in the dataset.
The horizontal axis means the user groups with a certain number of interactions. The left vertical axis is the value of the background histograms, which corresponds to the users' proportion in the dataset. The right vertical axis is the value of the polyline, which corresponds to $\sigma(\hat{y}_u)$. All the values are the average values of the users in the groups. As we can see, with the increase of the occurrence frequency of users in the dataset, the sigmoid scores of them also increase. This indicates that the user's activity is consistent with his/her conformity level. A similar phenomenon can be observed in Figure~\ref{diffyi} for different item groups. This shows our model's capability of capturing item popularity and users' conformity, thus benefiting the debias.

%% file: 02_relatedwork.tex
\section{related work}
In this section, we review existing work on Popularity Bias in Recommendation and Causal Inference in Recommendation, which are most relevant with this work.

\subsection{Popularity Bias in Recommendation}
Popularity bias is a common problem in recommender systems that popular items in the training dataset are frequently recommended. Researchers have explored many approaches \cite{abdollahpouri2017controlling,canamares2018should,jannach2015recommenders,canamares2017probabilistic,sun2019debiasing,enlighten193202,weifast,zheng2020dice} to analyzing and alleviating popularity bias in recommender systems. The first line of research is based on Inverse Propensity Weighting (IPW)~\cite{rosenbaum1983central} that is described in the above section. The core idea of this approach is reweighting the interactions in the training loss. For example, \citet{liang2016causal} propose to impose lower weights for popular items. Specifically, the weight is set as the inverse of item popularity. However, these previous methods ignore how popularity influence each specific interaction.

Another line of research tries to solve this problem through ranking adjustment. For instance, \citet{abdollahpouri2017controlling} propose a regularization-based approach that aims to improve the rank of long-tail items. \citet{abdollahpouri2019managing} introduce a re-ranking approach that can be applied to the output of the recommender systems. These approaches result in a trade-off between the recommendation accuracy and the coverage of unpopular items. They typically suffer from accuracy drop due to pushing the recommender to the long-tail in a brute manner. Unlike the existing work, we explore to eliminate popularity bias from a novel cause-effect perspective. We propose to capture the popularity bias through a multi-task training schema and remove the bias via counterfactual inference in the prediction stage.

\subsection{Causal Inference in Recommendation}

Causal inference is the science of systematically analyzing the relationship between a cause and its effect~\cite{pearl2009causality}. Recently, causal inference has gradually aroused people's attention and been exploited in a wide range of machine learning tasks, such as scene graph generation \cite{tang2020unbiased,chen2019counterfactual}, visual explanations \cite{martinez2019explaining}, vision-language multi-modal learning \cite{niu2020counterfactual,wang2020visual,qi2020two,zhang2020devlbert}, node classification~\cite{feng2020graph}, text classification~\cite{qian2021Counterfactual}, and natural language inference~\cite{Feng2021Empowering}. The main purpose of introducing causal inference in recommender systems is to remove the bias \cite{agarwal2019general, bellogin2017statistical,jiang2019degenerate, bottou2013counterfactual,schnabel2016recommendations,wang2020click,zhang2021causal}. We refer the readers to a systemic survey for more details \cite{chen2020bias}.

\paragraph{Inverse Propensity Weighting.} The first line of works is based on the Inverse Propensity Weighting (IPW). In \cite{liang2016causal}, the authors propose a framework consisted of two models: one exposure model and one preference model. Once the exposure model is estimated, the preference model is fit with weighted click data, where each click is weighted by the inverse of exposure estimated in the first model and thus be used to alleviate popularity bias. Some very similar models were proposed in \cite{schnabel2016recommendations, wang2018deconfounded}. 

\paragraph{Causality-oriented data.} The second line of works is working on leveraging additional debiased data. In \cite{bonner2018causal}, they propose to create an debiased training dataset as an auxiliary task to help the model trained in the skew dataset generalize better, which can also be used to relieve the popularity bias. They regard the large sample of the dataset as biased feedback data and model the recommendation as a domain adaption problem. But we argue that their method does not explicitly remove popularity bias and does not perform well on normal datasets. Noted that all these methods are aimed to reduce the user exposure bias. 

\paragraph{Causal embedding.} Another series of work is based on the probability, in \cite{liang2016modeling}, the authors present ExpoMF, a probabilistic approach for collaborative filtering on implicit data that directly incorporates user exposure to items into collaborative filtering. ExpoMF jointly models both users' exposure to an item, and their resulting click decisions, resulting in a model which naturally down-weights the expected, but ultimately un-clicked items. The exposure is modeled as a latent variable and the model infers its value from data. The popularity of items can be added as an exposure covariate and thus be used to alleviate popularity bias. This kind of works is based on probability and thus cannot be generalized to more prevalent settings. Moreover, they ignore how popularity influences each specific interaction. Similar to our work, \citet{zheng2020dice} also tries to mitigate popularity bias via causal approaches. 
The difference is that we analyze the causal relations in a fine-grained manner, consider the item popularity, user conformity and model their influence on recommendation. \cite{zheng2020dice} also lacks a systematic view of the mechanism of popularity bias.

%% file: 07_conclusion.tex
\section{conclusion and future work}
In this paper, we presented the first cause-effect view for alleviating popularity bias issue in recommender systems. We proposed the model-agnostic framework MACR which performs multi-task training according to the causal graph to assess the contribution of different causes on the ranking score. 
The counterfactual inference is performed to estimate the direct effect from item properties to the ranking score, which is removed to eliminate the popularity bias. Extensive experiments on five real-world recommendation datasets have demonstrated the effectiveness of MACR. 

This work represents one of the initial attempts to exploit causal reasoning for recommendation and opens up new research possibilities. In the future, we will extend our cause-effect look to more applications in recommender systems and explore other designs of the user and item module so as to better capture user conformity and item popularity. Moreover, we would like to explore how to incorporate various side information \cite{rendle2010factorization} and how our framework can be extended to alleviate other biases~\cite{chen2020bias} in recommender systems. In addition, we will study the simultaneous elimination of multiple types of biases such as popularity bias and exposure bias through counterfactual inference. Besides, we will explore the combination of causation and other relational domain knowledge~\cite{nie2020large}.


%% file: 08_appendix.tex
\appendix

\begin{figure*}[tb]
\centering
\subfigure[HR@K]{
\begin{minipage}[t]{0.3\textwidth}
\centering
\includegraphics[scale=0.21]{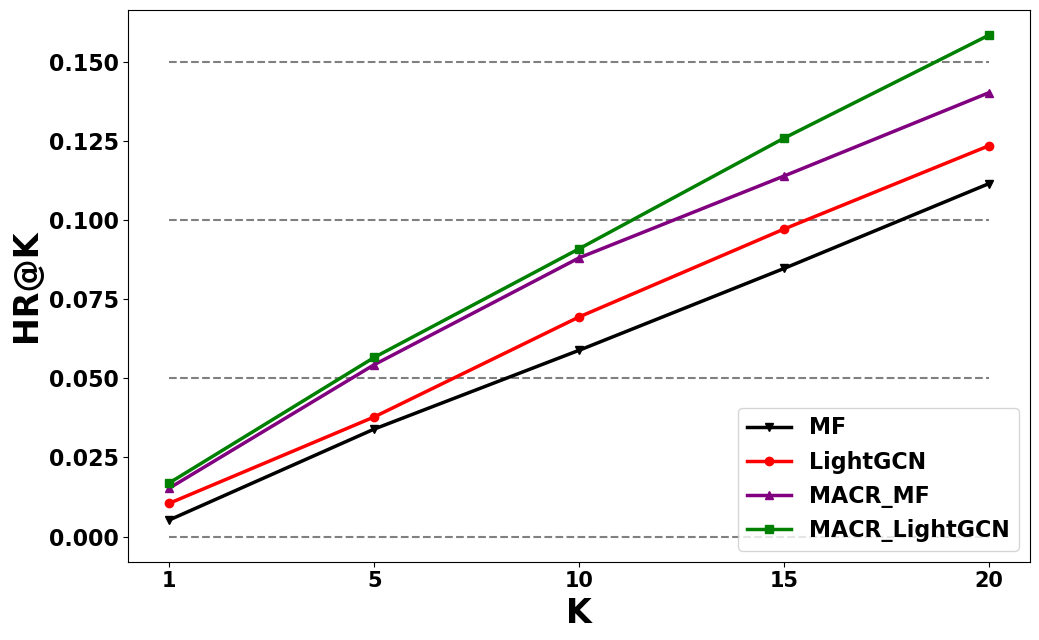}
\end{minipage}
}
\centering
\subfigure[NDCG@K]{ 
\begin{minipage}[t]{0.3\textwidth}
\centering
\includegraphics[scale=0.21]{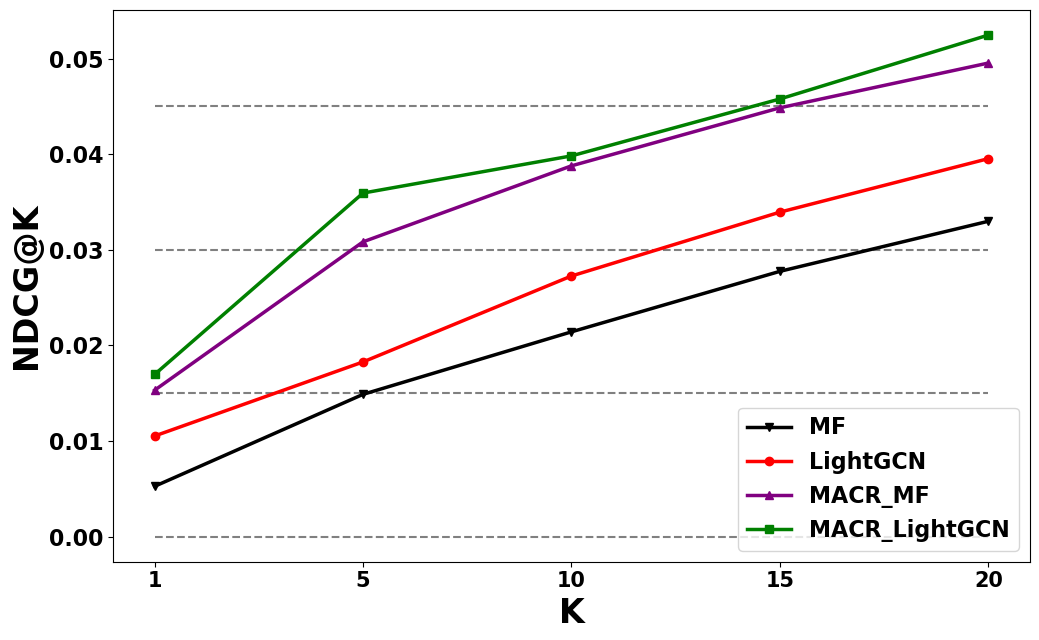}
\end{minipage}
}
\subfigure[Recall@K]{ 
\begin{minipage}[t]{0.3\textwidth}
\centering
\includegraphics[scale=0.21]{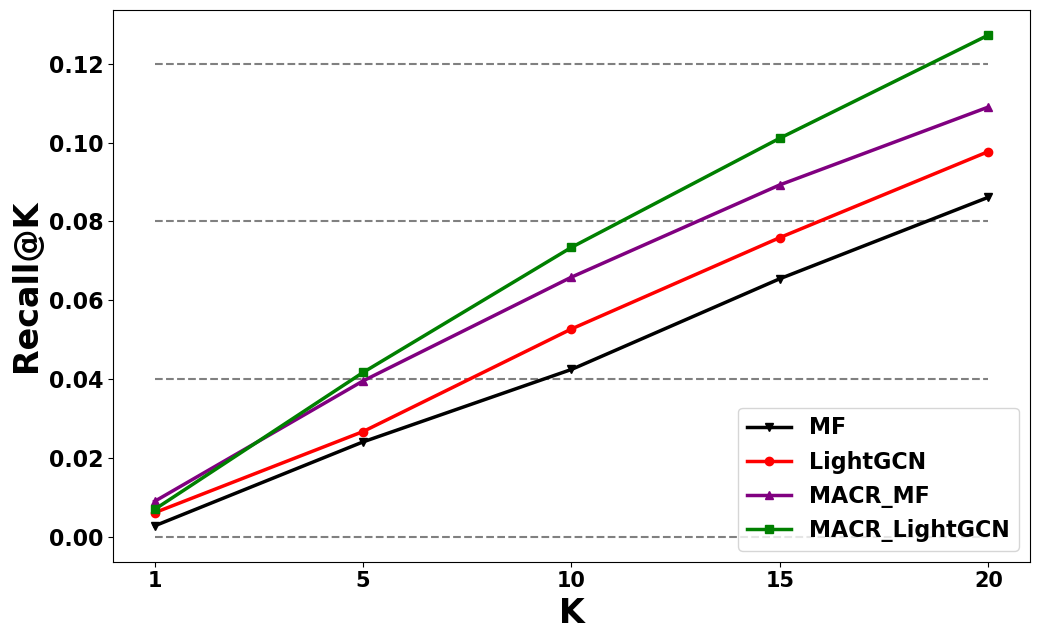}
\end{minipage}
}
\setlength{\abovecaptionskip}{-2pt}
\caption{Top-K recommendation performance on Adressa datasets w.r.t. HR@K, NDCG@K and Recall@K.}
\label{metric_k}
\vspace{-0.2cm}
\end{figure*}

\section{Inference Procedure}
\label{procudure}
Algorithm \ref{MACR} describes the procedure of our method and traditional recommendation system.

\begin{algorithm}[htbp]
\caption{Inference}
\label{MACR}
\begin{flushleft}
\textbf{Input:} Backbone recommender $Y_k$, Item module $Y_i$, User module $Y_u$, User $u$, Item $i$, Reference status $c$. \\
\textbf{Output:} $\hat{y}_{ui}$ \\
\end{flushleft}
\begin{algorithmic}[1]
\STATE/*  Model Agnostic Counterfactual Reasoning  */
\STATE
$\hat{y}_k=Y_k(K(u,i))$;
\STATE
$
\hat{y}_i=Y_i(i)
$;
\STATE
$
\hat{y}_u=Y_i(u)
$;
\IF{$Is\_Training$}
\STATE
$\hat{y}_{ui} = \hat{y}_{k}*\sigma(\hat{y}_{i})*\sigma(\hat{y}_{u})$;
\ELSE
\STATE
$\hat{y}_{ui} = \hat{y}_{k}*\sigma(\hat{y}_{i})*\sigma(\hat{y}_{u})-c*\sigma(\hat{y}_{i})*\sigma(\hat{y}_{u})$;
\ENDIF
\STATE/*  Traditional Recommender  */
\STATE
$\hat{y}_{ui}=Y_k(K(u,i))$;
\end{algorithmic}
\end{algorithm}

\section{Implementation details}
\label{parametersetting}
We implement MACR in Tensorflow  \cite{abadi2016tensorflow}. The embedding size is fixed to 64 for all models and the embedding parameters are initialized with the Xavier method \cite{glorot2010understanding}. We optimize all models with Adam \cite{kingma2014adam} except for ExpoMF which is trained in a probabilistic manner as per the original paper \cite{liang2016modeling}. For all methods, we use the default learning rate of 0.001 and default mini-batch size of 1024 (on ML10M and Globo, we increase the mini-batch size to 8192 to speed up training). Also, we choose binarized cross-entropy loss for all models for a fair comparison. For the LightGCN model, we utilize two layers of graph convolution network to obtain the best results. For the Reg model, the coefficient for the item-based regularization is set to 1e-4 because it works best. For DICE, we keep all the optimal setting in their paper except replacing the regularization term  $L_{discrepancy}$ from $dCor$ with another option - $L2$. Because with our large-scale dataset, computing $dCor$ will be out of memory for the 2080Ti GPU. It is also suggested in their paper. For ExpoMF, the initial value of $\mu$ is tuned in the range of $\{0.1, 0.05, 0.01, 0.005, 0.001\}$ as suggested by the author. For CausE, as their model training needs one biased dataset and another debiased dataset, we split 10\% of the train data as we mentioned in Section \ref{evaluation} to build an additional debiased dataset for it. For our MACR\_MF and MACR\_LightGCN, the trade-off parameters $\alpha$ and $\beta$ in Eq.~\eqref{ab} are both searched in the range of $\{1e-5,1e-4,1e-3,1e-2\}$ and set to 1e-3 by default. The $c$ in Eq.~\eqref{eq:ranking_test} is tuned in the range of $\{20, 22, ..., 40\}$. The number of training epochs is fixed to 1000. The L2 regularization coefficient is set to 1e-5 by default. 

\section{Supplementary experiments}
\subsection{Metrics with different Ks}
\label{metric}
Figure \ref{metric_k} reports our experimental results on Adressa dataset w.r.t. HR@K, NDCG@K and Recall@K where $K=\{1,5,10,15,20\}$. It shows the effectiveness of MACR which can improve MF and LightGCN on different metrics with a large margin. Due to space limitation, we show the results on the Adressa dataset only, and the results on the other four datasets show the same trend.

\begin{table}[]
\caption{Effect of $\alpha$ on MACR\_MF.}
\vspace{-0.3cm}
\begin{tabular}{|c|l|l|l|}
\hline
\multicolumn{1}{|l|}{} & \multicolumn{1}{c|}{HR@20} & \multicolumn{1}{c|}{Recall@20} & \multicolumn{1}{c|}{NDCG@20} \\ \hline
1e-5                      & 0.133                    & 0.104                        & 0.045                      \\ \hline
1e-4                   & 0.139                    & 0.108                        & 0.049                      \\ \hline
1e-3                   & \textbf{0.140}                    & \textbf{0.109}                        & \textbf{0.050}                      \\ \hline
1e-2                   & 0.137                    & 0.108                        & 0.048                      \\ \hline

\end{tabular}
\label{alphaeff}
\vspace{-0.3cm}
\end{table}

\begin{table}[]
\caption{Effect of $\beta$ on MACR\_MF.}
\vspace{-0.3cm}
\begin{tabular}{|c|l|l|l|}
\hline
\multicolumn{1}{|l|}{} & \multicolumn{1}{c|}{HR@20} & \multicolumn{1}{c|}{Recall@20} & \multicolumn{1}{c|}{NDCG@20} \\ \hline
1e-5                      & 0.139                    & 0.108                        & 0.049                      \\ \hline
1e-4                   & 0.139                    & 0.109                        & 0.049                      \\ \hline
1e-3                   & \textbf{0.140}           & \textbf{0.109}               & \textbf{0.050}             \\ \hline
1e-2                   & 0.139                    & 0.108                        & 0.049                      \\ \hline
\end{tabular}
\label{betaeff}
\vspace{-0.3cm}
\end{table}

\subsection{Effect of hyper-parameters}
\label{hyper}
As formulated in the loss function Eq.~\eqref{ab}, $\alpha$ is the trade-off hyper-parameter which balances the contribution of the recommendation model loss and the item model loss while $\beta$ is to balance the recommendation model loss and the user loss. To investigate the benefit of item loss and user loss, we conduct experiments of MACR\_MF on the typical Adressa dataset with varying $\alpha$ and $\beta$ respectively. In particular, we search their values in the range of \{1e-5, 1e-4, 1e-3, 1e-2\}. When varying one parameter, the other is set as constant 1e-3. From Table~\ref{alphaeff} and Table~\ref{betaeff} we have the following findings:
\begin{itemize}[leftmargin=*]
    \item As $\alpha$ increases from 1e-5 to 1e-3, the performance of MACR will become better. This result indicates the importance of capturing item popularity bias. A similar trend can be observed by varying $\beta$ from 1e-5 to 1e-3 and it demonstrates the benefit of capturing users' conformity.
    \item However, when $\alpha$ or $\beta$ surpasses a threshold (1e-3), the performance becomes worse with a further increase of the parameters. As parameters become further larger, the training of the recommendation model will be less important, which brings the worse results.
\end{itemize}

%% file: main.bbl

\begin{thebibliography}{60}


\ifx \showCODEN    \undefined \def \showCODEN     #1{\unskip}     \fi
\ifx \showDOI      \undefined \def \showDOI       #1{#1}\fi
\ifx \showISBNx    \undefined \def \showISBNx     #1{\unskip}     \fi
\ifx \showISBNxiii \undefined \def \showISBNxiii  #1{\unskip}     \fi
\ifx \showISSN     \undefined \def \showISSN      #1{\unskip}     \fi
\ifx \showLCCN     \undefined \def \showLCCN      #1{\unskip}     \fi
\ifx \shownote     \undefined \def \shownote      #1{#1}          \fi
\ifx \showarticletitle \undefined \def \showarticletitle #1{#1}   \fi
\ifx \showURL      \undefined \def \showURL       {\relax}        \fi
\providecommand\bibfield[2]{#2}
\providecommand\bibinfo[2]{#2}
\providecommand\natexlab[1]{#1}
\providecommand\showeprint[2][]{arXiv:#2}

\bibitem[\protect\citeauthoryear{Abadi, Barham, Chen, Chen, Davis, Dean, Devin,
  Ghemawat, Irving, Isard, et~al\mbox{.}}{Abadi et~al\mbox{.}}{2016}]%
        {abadi2016tensorflow}
\bibfield{author}{\bibinfo{person}{Mart{\'\i}n Abadi}, \bibinfo{person}{Paul
  Barham}, \bibinfo{person}{Jianmin Chen}, \bibinfo{person}{Zhifeng Chen},
  \bibinfo{person}{Andy Davis}, \bibinfo{person}{Jeffrey Dean},
  \bibinfo{person}{Matthieu Devin}, \bibinfo{person}{Sanjay Ghemawat},
  \bibinfo{person}{Geoffrey Irving}, \bibinfo{person}{Michael Isard},
  {et~al\mbox{.}}} \bibinfo{year}{2016}\natexlab{}.
\newblock \showarticletitle{Tensorflow: A system for large-scale machine
  learning}. In \bibinfo{booktitle}{\emph{OSDI}}. \bibinfo{pages}{265–283}.
\newblock


\bibitem[\protect\citeauthoryear{Abdollahpouri, Burke, and
  Mobasher}{Abdollahpouri et~al\mbox{.}}{2017}]%
        {abdollahpouri2017controlling}
\bibfield{author}{\bibinfo{person}{Himan Abdollahpouri}, \bibinfo{person}{Robin
  Burke}, {and} \bibinfo{person}{Bamshad Mobasher}.}
  \bibinfo{year}{2017}\natexlab{}.
\newblock \showarticletitle{Controlling popularity bias in learning-to-rank
  recommendation}. In \bibinfo{booktitle}{\emph{RecSys}}.
  \bibinfo{pages}{42--46}.
\newblock


\bibitem[\protect\citeauthoryear{Abdollahpouri, Burke, and
  Mobasher}{Abdollahpouri et~al\mbox{.}}{2019}]%
        {abdollahpouri2019managing}
\bibfield{author}{\bibinfo{person}{Himan Abdollahpouri}, \bibinfo{person}{Robin
  Burke}, {and} \bibinfo{person}{Bamshad Mobasher}.}
  \bibinfo{year}{2019}\natexlab{}.
\newblock \showarticletitle{Managing popularity bias in recommender systems
  with personalized re-ranking}. In \bibinfo{booktitle}{\emph{FLAIRS}}.
\newblock


\bibitem[\protect\citeauthoryear{Agarwal, Takatsu, Zaitsev, and
  Joachims}{Agarwal et~al\mbox{.}}{2019}]%
        {agarwal2019general}
\bibfield{author}{\bibinfo{person}{Aman Agarwal}, \bibinfo{person}{Kenta
  Takatsu}, \bibinfo{person}{Ivan Zaitsev}, {and} \bibinfo{person}{Thorsten
  Joachims}.} \bibinfo{year}{2019}\natexlab{}.
\newblock \showarticletitle{A general framework for counterfactual
  learning-to-rank}. In \bibinfo{booktitle}{\emph{SIGIR}}.
  \bibinfo{pages}{5--14}.
\newblock


\bibitem[\protect\citeauthoryear{Bellog{\'\i}n, Castells, and
  Cantador}{Bellog{\'\i}n et~al\mbox{.}}{2017}]%
        {bellogin2017statistical}
\bibfield{author}{\bibinfo{person}{Alejandro Bellog{\'\i}n},
  \bibinfo{person}{Pablo Castells}, {and} \bibinfo{person}{Iv{\'a}n Cantador}.}
  \bibinfo{year}{2017}\natexlab{}.
\newblock \showarticletitle{Statistical biases in Information Retrieval metrics
  for recommender systems}.
\newblock \bibinfo{journal}{\emph{Inf. Retr. J.}} (\bibinfo{year}{2017}),
  \bibinfo{pages}{606--634}.
\newblock


\bibitem[\protect\citeauthoryear{Berg, Kipf, and Welling}{Berg
  et~al\mbox{.}}{2018}]%
        {berg2017graph}
\bibfield{author}{\bibinfo{person}{Rianne van~den Berg},
  \bibinfo{person}{Thomas~N Kipf}, {and} \bibinfo{person}{Max Welling}.}
  \bibinfo{year}{2018}\natexlab{}.
\newblock \showarticletitle{Graph convolutional matrix completion}.
\newblock \bibinfo{journal}{\emph{KDD Deep Learning Day}}
  (\bibinfo{year}{2018}).
\newblock


\bibitem[\protect\citeauthoryear{Bonner and Vasile}{Bonner and Vasile}{2018}]%
        {bonner2018causal}
\bibfield{author}{\bibinfo{person}{Stephen Bonner} {and}
  \bibinfo{person}{Flavian Vasile}.} \bibinfo{year}{2018}\natexlab{}.
\newblock \showarticletitle{Causal embeddings for recommendation}. In
  \bibinfo{booktitle}{\emph{RecSys}}. \bibinfo{pages}{104--112}.
\newblock


\bibitem[\protect\citeauthoryear{Bottou, Peters, Qui{\~n}onero-Candela,
  Charles, Chickering, Portugaly, Ray, Simard, and Snelson}{Bottou
  et~al\mbox{.}}{2013}]%
        {bottou2013counterfactual}
\bibfield{author}{\bibinfo{person}{L{\'e}on Bottou}, \bibinfo{person}{Jonas
  Peters}, \bibinfo{person}{Joaquin Qui{\~n}onero-Candela},
  \bibinfo{person}{Denis~X Charles}, \bibinfo{person}{D~Max Chickering},
  \bibinfo{person}{Elon Portugaly}, \bibinfo{person}{Dipankar Ray},
  \bibinfo{person}{Patrice Simard}, {and} \bibinfo{person}{Ed Snelson}.}
  \bibinfo{year}{2013}\natexlab{}.
\newblock \showarticletitle{Counterfactual reasoning and learning systems: The
  example of computational advertising}.
\newblock \bibinfo{journal}{\emph{JMLR}} (\bibinfo{year}{2013}),
  \bibinfo{pages}{3207--3260}.
\newblock


\bibitem[\protect\citeauthoryear{Ca{\~n}amares and Castells}{Ca{\~n}amares and
  Castells}{2017}]%
        {canamares2017probabilistic}
\bibfield{author}{\bibinfo{person}{Roc{\'\i}o Ca{\~n}amares} {and}
  \bibinfo{person}{Pablo Castells}.} \bibinfo{year}{2017}\natexlab{}.
\newblock \showarticletitle{A probabilistic reformulation of memory-based
  collaborative filtering: Implications on popularity biases}. In
  \bibinfo{booktitle}{\emph{SIGIR}}. \bibinfo{pages}{215--224}.
\newblock


\bibitem[\protect\citeauthoryear{Ca{\~n}amares and Castells}{Ca{\~n}amares and
  Castells}{2018}]%
        {canamares2018should}
\bibfield{author}{\bibinfo{person}{Roc{\'\i}o Ca{\~n}amares} {and}
  \bibinfo{person}{Pablo Castells}.} \bibinfo{year}{2018}\natexlab{}.
\newblock \showarticletitle{Should I follow the crowd? A probabilistic analysis
  of the effectiveness of popularity in recommender systems}. In
  \bibinfo{booktitle}{\emph{SIGIR}}. \bibinfo{pages}{415--424}.
\newblock


\bibitem[\protect\citeauthoryear{Chaney, Stewart, and Engelhardt}{Chaney
  et~al\mbox{.}}{2018}]%
        {algconfound}
\bibfield{author}{\bibinfo{person}{Allison J.~B. Chaney},
  \bibinfo{person}{Brandon~M. Stewart}, {and} \bibinfo{person}{Barbara~E.
  Engelhardt}.} \bibinfo{year}{2018}\natexlab{}.
\newblock \showarticletitle{How Algorithmic Confounding in Recommendation
  Systems Increases Homogeneity and Decreases Utility}. In
  \bibinfo{booktitle}{\emph{RecSys}}. \bibinfo{pages}{224–232}.
\newblock
\showISBNx{9781450359016}


\bibitem[\protect\citeauthoryear{Chen, Dong, Wang, Feng, Wang, and He}{Chen
  et~al\mbox{.}}{2020}]%
        {chen2020bias}
\bibfield{author}{\bibinfo{person}{Jiawei Chen}, \bibinfo{person}{Hande Dong},
  \bibinfo{person}{Xiang Wang}, \bibinfo{person}{Fuli Feng},
  \bibinfo{person}{Meng Wang}, {and} \bibinfo{person}{Xiangnan He}.}
  \bibinfo{year}{2020}\natexlab{}.
\newblock \showarticletitle{Bias and Debias in Recommender System: A Survey and
  Future Directions}.
\newblock \bibinfo{journal}{\emph{arXiv preprint arXiv:2010.03240}}
  (\bibinfo{year}{2020}).
\newblock


\bibitem[\protect\citeauthoryear{Chen, Zhang, Xiao, He, Pu, and Chang}{Chen
  et~al\mbox{.}}{2019}]%
        {chen2019counterfactual}
\bibfield{author}{\bibinfo{person}{Long Chen}, \bibinfo{person}{Hanwang Zhang},
  \bibinfo{person}{Jun Xiao}, \bibinfo{person}{Xiangnan He},
  \bibinfo{person}{Shiliang Pu}, {and} \bibinfo{person}{Shih-Fu Chang}.}
  \bibinfo{year}{2019}\natexlab{}.
\newblock \showarticletitle{Counterfactual critic multi-agent training for
  scene graph generation}. In \bibinfo{booktitle}{\emph{ICCV}}.
  \bibinfo{pages}{4613--4623}.
\newblock


\bibitem[\protect\citeauthoryear{de~Souza Pereira~Moreira, Ferreira, and
  da~Cunha}{de~Souza Pereira~Moreira et~al\mbox{.}}{2018}]%
        {gabriel2019contextual}
\bibfield{author}{\bibinfo{person}{Gabriel de Souza Pereira~Moreira},
  \bibinfo{person}{Felipe Ferreira}, {and} \bibinfo{person}{Adilson~Marques da
  Cunha}.} \bibinfo{year}{2018}\natexlab{}.
\newblock \showarticletitle{News session-based recommendations using deep
  neural networks}. In \bibinfo{booktitle}{\emph{Workshop on Deep Learning at
  RecSys}}. \bibinfo{pages}{15--23}.
\newblock


\bibitem[\protect\citeauthoryear{Feng, Huang, He, Xin, Wang, and Chua}{Feng
  et~al\mbox{.}}{2021a}]%
        {feng2020graph}
\bibfield{author}{\bibinfo{person}{Fuli Feng}, \bibinfo{person}{Weiran Huang},
  \bibinfo{person}{Xiangnan He}, \bibinfo{person}{Xin Xin},
  \bibinfo{person}{Qifan Wang}, {and} \bibinfo{person}{Tat-Seng~Chua Chua}.}
  \bibinfo{year}{2021}\natexlab{a}.
\newblock \showarticletitle{Should Graph Convolution Trust Neighbors? A Simple
  Causal Inference Method}. In \bibinfo{booktitle}{\emph{SIGIR}}.
\newblock


\bibitem[\protect\citeauthoryear{Feng, Zhang, He, Zhang, and Chua}{Feng
  et~al\mbox{.}}{2021b}]%
        {Feng2021Empowering}
\bibfield{author}{\bibinfo{person}{Fuli Feng}, \bibinfo{person}{Jizhi Zhang},
  \bibinfo{person}{Xiangnan He}, \bibinfo{person}{Hanwang Zhang}, {and}
  \bibinfo{person}{Tat-Seng Chua}.} \bibinfo{year}{2021}\natexlab{b}.
\newblock \showarticletitle{Empowering Language Understanding with
  Counterfactual Reasoning}. In \bibinfo{booktitle}{\emph{ACL-IJCNLP
  Findings}}.
\newblock


\bibitem[\protect\citeauthoryear{Glorot and Bengio}{Glorot and Bengio}{2010}]%
        {glorot2010understanding}
\bibfield{author}{\bibinfo{person}{Xavier Glorot} {and} \bibinfo{person}{Yoshua
  Bengio}.} \bibinfo{year}{2010}\natexlab{}.
\newblock \showarticletitle{Understanding the difficulty of training deep
  feedforward neural networks}. In \bibinfo{booktitle}{\emph{AISTATS}}.
  \bibinfo{pages}{249--256}.
\newblock


\bibitem[\protect\citeauthoryear{Gulla, Zhang, Liu, {\"O}zg{\"o}bek, and
  Su}{Gulla et~al\mbox{.}}{2017}]%
        {gulla2017adressa}
\bibfield{author}{\bibinfo{person}{Jon~Atle Gulla}, \bibinfo{person}{Lemei
  Zhang}, \bibinfo{person}{Peng Liu}, \bibinfo{person}{{\"O}zlem
  {\"O}zg{\"o}bek}, {and} \bibinfo{person}{Xiaomeng Su}.}
  \bibinfo{year}{2017}\natexlab{}.
\newblock \showarticletitle{The Adressa dataset for news recommendation}. In
  \bibinfo{booktitle}{\emph{WI}}. \bibinfo{pages}{1042--1048}.
\newblock


\bibitem[\protect\citeauthoryear{Guo, Zhang, and Yorke-Smith}{Guo
  et~al\mbox{.}}{2015}]%
        {guo2015trustsvd}
\bibfield{author}{\bibinfo{person}{Guibing Guo}, \bibinfo{person}{Jie Zhang},
  {and} \bibinfo{person}{Neil Yorke-Smith}.} \bibinfo{year}{2015}\natexlab{}.
\newblock \showarticletitle{Trustsvd: Collaborative filtering with both the
  explicit and implicit influence of user trust and of item ratings.}. In
  \bibinfo{booktitle}{\emph{AAAI}}. \bibinfo{pages}{123--125}.
\newblock


\bibitem[\protect\citeauthoryear{He, Deng, Wang, Li, Zhang, and Wang}{He
  et~al\mbox{.}}{2020}]%
        {he2020lightgcn}
\bibfield{author}{\bibinfo{person}{Xiangnan He}, \bibinfo{person}{Kuan Deng},
  \bibinfo{person}{Xiang Wang}, \bibinfo{person}{Yan Li},
  \bibinfo{person}{Yongdong Zhang}, {and} \bibinfo{person}{Meng Wang}.}
  \bibinfo{year}{2020}\natexlab{}.
\newblock \showarticletitle{LightGCN: Simplifying and Powering Graph
  Convolution Network for Recommendation}. In
  \bibinfo{booktitle}{\emph{SIGIR}}. \bibinfo{pages}{639–648}.
\newblock


\bibitem[\protect\citeauthoryear{He, Du, Wang, Tian, Tang, and Chua}{He
  et~al\mbox{.}}{2018}]%
        {he2018outer}
\bibfield{author}{\bibinfo{person}{Xiangnan He}, \bibinfo{person}{Xiaoyu Du},
  \bibinfo{person}{Xiang Wang}, \bibinfo{person}{Feng Tian},
  \bibinfo{person}{Jinhui Tang}, {and} \bibinfo{person}{Tat-Seng Chua}.}
  \bibinfo{year}{2018}\natexlab{}.
\newblock \showarticletitle{Outer product-based neural collaborative
  filtering}. In \bibinfo{booktitle}{\emph{IJCAI}}.
  \bibinfo{pages}{2227--2233}.
\newblock


\bibitem[\protect\citeauthoryear{He, Liao, Zhang, Nie, Hu, and Chua}{He
  et~al\mbox{.}}{2017}]%
        {he2017neural}
\bibfield{author}{\bibinfo{person}{Xiangnan He}, \bibinfo{person}{Lizi Liao},
  \bibinfo{person}{Hanwang Zhang}, \bibinfo{person}{Liqiang Nie},
  \bibinfo{person}{Xia Hu}, {and} \bibinfo{person}{Tat-Seng Chua}.}
  \bibinfo{year}{2017}\natexlab{}.
\newblock \showarticletitle{Neural collaborative filtering}. In
  \bibinfo{booktitle}{\emph{WWW}}. \bibinfo{pages}{173--182}.
\newblock


\bibitem[\protect\citeauthoryear{Jadidinejad, Macdonald, and Ounis}{Jadidinejad
  et~al\mbox{.}}{2019}]%
        {enlighten193202}
\bibfield{author}{\bibinfo{person}{Amir Jadidinejad}, \bibinfo{person}{Craig
  Macdonald}, {and} \bibinfo{person}{Iadh Ounis}.}
  \bibinfo{year}{2019}\natexlab{}.
\newblock \showarticletitle{How Sensitive is Recommendation Systems' Offline
  Evaluation to Popularity?}. In \bibinfo{booktitle}{\emph{REVEAL Workshop at
  RecSys}}.
\newblock


\bibitem[\protect\citeauthoryear{Jannach, Lerche, Kamehkhosh, and
  Jugovac}{Jannach et~al\mbox{.}}{2015}]%
        {jannach2015recommenders}
\bibfield{author}{\bibinfo{person}{Dietmar Jannach}, \bibinfo{person}{Lukas
  Lerche}, \bibinfo{person}{Iman Kamehkhosh}, {and} \bibinfo{person}{Michael
  Jugovac}.} \bibinfo{year}{2015}\natexlab{}.
\newblock \showarticletitle{What recommenders recommend: an analysis of
  recommendation biases and possible countermeasures}.
\newblock \bibinfo{journal}{\emph{User Model User-adapt Interact}}
  (\bibinfo{year}{2015}), \bibinfo{pages}{427--491}.
\newblock


\bibitem[\protect\citeauthoryear{Jiang, Chiappa, Lattimore, Gy{\"o}rgy, and
  Kohli}{Jiang et~al\mbox{.}}{2019}]%
        {jiang2019degenerate}
\bibfield{author}{\bibinfo{person}{Ray Jiang}, \bibinfo{person}{Silvia
  Chiappa}, \bibinfo{person}{Tor Lattimore}, \bibinfo{person}{Andr{\'a}s
  Gy{\"o}rgy}, {and} \bibinfo{person}{Pushmeet Kohli}.}
  \bibinfo{year}{2019}\natexlab{}.
\newblock \showarticletitle{Degenerate feedback loops in recommender systems}.
  In \bibinfo{booktitle}{\emph{AIES}}. \bibinfo{pages}{383--390}.
\newblock


\bibitem[\protect\citeauthoryear{Kabbur, Ning, and Karypis}{Kabbur
  et~al\mbox{.}}{2013}]%
        {kabbur2013fism}
\bibfield{author}{\bibinfo{person}{Santosh Kabbur}, \bibinfo{person}{Xia Ning},
  {and} \bibinfo{person}{George Karypis}.} \bibinfo{year}{2013}\natexlab{}.
\newblock \showarticletitle{Fism: factored item similarity models for top-n
  recommender systems}. In \bibinfo{booktitle}{\emph{KDD}}.
  \bibinfo{pages}{659--667}.
\newblock


\bibitem[\protect\citeauthoryear{Kingma and Ba}{Kingma and Ba}{2014}]%
        {kingma2014adam}
\bibfield{author}{\bibinfo{person}{Diederik~P Kingma} {and}
  \bibinfo{person}{Jimmy Ba}.} \bibinfo{year}{2014}\natexlab{}.
\newblock \showarticletitle{Adam: A method for stochastic optimization}. In
  \bibinfo{booktitle}{\emph{ICLR}}.
\newblock


\bibitem[\protect\citeauthoryear{Koren, Bell, and Volinsky}{Koren
  et~al\mbox{.}}{2009}]%
        {koren2009matrix}
\bibfield{author}{\bibinfo{person}{Yehuda Koren}, \bibinfo{person}{Robert
  Bell}, {and} \bibinfo{person}{Chris Volinsky}.}
  \bibinfo{year}{2009}\natexlab{}.
\newblock \showarticletitle{Matrix factorization techniques for recommender
  systems}.
\newblock \bibinfo{journal}{\emph{Computer}} (\bibinfo{year}{2009}),
  \bibinfo{pages}{30--37}.
\newblock


\bibitem[\protect\citeauthoryear{Lei, He, Miao, Wu, Hong, Kan, and Chua}{Lei
  et~al\mbox{.}}{2020}]%
        {lei2020estimation}
\bibfield{author}{\bibinfo{person}{Wenqiang Lei}, \bibinfo{person}{Xiangnan
  He}, \bibinfo{person}{Yisong Miao}, \bibinfo{person}{Qingyun Wu},
  \bibinfo{person}{Richang Hong}, \bibinfo{person}{Min-Yen Kan}, {and}
  \bibinfo{person}{Tat-Seng Chua}.} \bibinfo{year}{2020}\natexlab{}.
\newblock \showarticletitle{Estimation-action-reflection: Towards deep
  interaction between conversational and recommender systems}. In
  \bibinfo{booktitle}{\emph{WSDM}}. \bibinfo{pages}{304--312}.
\newblock


\bibitem[\protect\citeauthoryear{Liang, Charlin, and Blei}{Liang
  et~al\mbox{.}}{2016a}]%
        {liang2016causal}
\bibfield{author}{\bibinfo{person}{Dawen Liang}, \bibinfo{person}{Laurent
  Charlin}, {and} \bibinfo{person}{David~M Blei}.}
  \bibinfo{year}{2016}\natexlab{a}.
\newblock \showarticletitle{Causal inference for recommendation}. In
  \bibinfo{booktitle}{\emph{Workshop at UAI}}.
\newblock


\bibitem[\protect\citeauthoryear{Liang, Charlin, McInerney, and Blei}{Liang
  et~al\mbox{.}}{2016b}]%
        {liang2016modeling}
\bibfield{author}{\bibinfo{person}{Dawen Liang}, \bibinfo{person}{Laurent
  Charlin}, \bibinfo{person}{James McInerney}, {and} \bibinfo{person}{David~M
  Blei}.} \bibinfo{year}{2016}\natexlab{b}.
\newblock \showarticletitle{Modeling user exposure in recommendation}. In
  \bibinfo{booktitle}{\emph{WWW}}. \bibinfo{pages}{951--961}.
\newblock


\bibitem[\protect\citeauthoryear{Mart{\'\i}nez and Marca}{Mart{\'\i}nez and
  Marca}{2019}]%
        {martinez2019explaining}
\bibfield{author}{\bibinfo{person}{{\'A}lvaro~Parafita Mart{\'\i}nez} {and}
  \bibinfo{person}{Jordi~Vitri{\`a} Marca}.} \bibinfo{year}{2019}\natexlab{}.
\newblock \showarticletitle{Explaining Visual Models by Causal Attribution}. In
  \bibinfo{booktitle}{\emph{ICCV Workshop}}. \bibinfo{pages}{4167--4175}.
\newblock


\bibitem[\protect\citeauthoryear{Nie, Li, Feng, Song, Wang, and Wang}{Nie
  et~al\mbox{.}}{2020}]%
        {nie2020large}
\bibfield{author}{\bibinfo{person}{Liqiang Nie}, \bibinfo{person}{Yongqi Li},
  \bibinfo{person}{Fuli Feng}, \bibinfo{person}{Xuemeng Song},
  \bibinfo{person}{Meng Wang}, {and} \bibinfo{person}{Yinglong Wang}.}
  \bibinfo{year}{2020}\natexlab{}.
\newblock \showarticletitle{Large-scale question tagging via joint
  question-topic embedding learning}.
\newblock \bibinfo{journal}{\emph{ACM TOIS}} \bibinfo{volume}{38},
  \bibinfo{number}{2} (\bibinfo{year}{2020}), \bibinfo{pages}{1--23}.
\newblock


\bibitem[\protect\citeauthoryear{Niu, Tang, Zhang, Lu, Hua, and Wen}{Niu
  et~al\mbox{.}}{2021}]%
        {niu2020counterfactual}
\bibfield{author}{\bibinfo{person}{Yulei Niu}, \bibinfo{person}{Kaihua Tang},
  \bibinfo{person}{Hanwang Zhang}, \bibinfo{person}{Zhiwu Lu},
  \bibinfo{person}{Xian-Sheng Hua}, {and} \bibinfo{person}{Ji-Rong Wen}.}
  \bibinfo{year}{2021}\natexlab{}.
\newblock \showarticletitle{Counterfactual VQA: A Cause-Effect Look at Language
  Bias}. In \bibinfo{booktitle}{\emph{CVPR}}.
\newblock


\bibitem[\protect\citeauthoryear{Pearl}{Pearl}{2009}]%
        {pearl2009causality}
\bibfield{author}{\bibinfo{person}{Judea Pearl}.}
  \bibinfo{year}{2009}\natexlab{}.
\newblock \bibinfo{booktitle}{\emph{Causality}}.
\newblock \bibinfo{publisher}{Cambridge Uiversity Press}.
\newblock


\bibitem[\protect\citeauthoryear{Perc}{Perc}{2014}]%
        {perc2014matthew}
\bibfield{author}{\bibinfo{person}{Matja{\v{z}} Perc}.}
  \bibinfo{year}{2014}\natexlab{}.
\newblock \showarticletitle{The Matthew effect in empirical data}.
\newblock \bibinfo{journal}{\emph{J R Soc Interface}} (\bibinfo{year}{2014}),
  \bibinfo{pages}{20140378}.
\newblock


\bibitem[\protect\citeauthoryear{Qi, Niu, Huang, and Zhang}{Qi
  et~al\mbox{.}}{2020}]%
        {qi2020two}
\bibfield{author}{\bibinfo{person}{Jiaxin Qi}, \bibinfo{person}{Yulei Niu},
  \bibinfo{person}{Jianqiang Huang}, {and} \bibinfo{person}{Hanwang Zhang}.}
  \bibinfo{year}{2020}\natexlab{}.
\newblock \showarticletitle{Two causal principles for improving visual dialog}.
  In \bibinfo{booktitle}{\emph{CVPR}}. \bibinfo{pages}{10860--10869}.
\newblock


\bibitem[\protect\citeauthoryear{Qian, Feng, Wen, Ma, and Xie}{Qian
  et~al\mbox{.}}{2021}]%
        {qian2021Counterfactual}
\bibfield{author}{\bibinfo{person}{Chen Qian}, \bibinfo{person}{Fuli Feng},
  \bibinfo{person}{Lijie Wen}, \bibinfo{person}{Chunping Ma}, {and}
  \bibinfo{person}{Pengjun Xie}.} \bibinfo{year}{2021}\natexlab{}.
\newblock \showarticletitle{Counterfactual Inference for Text Classification
  Debiasing}. In \bibinfo{booktitle}{\emph{ACL-IJCNLP}}.
\newblock


\bibitem[\protect\citeauthoryear{Rendle}{Rendle}{2010}]%
        {rendle2010factorization}
\bibfield{author}{\bibinfo{person}{Steffen Rendle}.}
  \bibinfo{year}{2010}\natexlab{}.
\newblock \showarticletitle{Factorization machines}. In
  \bibinfo{booktitle}{\emph{ICDM}}. \bibinfo{pages}{995--1000}.
\newblock


\bibitem[\protect\citeauthoryear{Rendle, Zhang, and Koren}{Rendle
  et~al\mbox{.}}{2019}]%
        {rendle2019difficulty}
\bibfield{author}{\bibinfo{person}{Steffen Rendle}, \bibinfo{person}{Li Zhang},
  {and} \bibinfo{person}{Yehuda Koren}.} \bibinfo{year}{2019}\natexlab{}.
\newblock \showarticletitle{On the difficulty of evaluating baselines: A study
  on recommender systems}.
\newblock \bibinfo{journal}{\emph{arXiv preprint arXiv:1905.01395}}
  (\bibinfo{year}{2019}).
\newblock


\bibitem[\protect\citeauthoryear{Rosenbaum and Rubin}{Rosenbaum and
  Rubin}{1983}]%
        {rosenbaum1983central}
\bibfield{author}{\bibinfo{person}{Paul~R Rosenbaum} {and}
  \bibinfo{person}{Donald~B Rubin}.} \bibinfo{year}{1983}\natexlab{}.
\newblock \showarticletitle{The central role of the propensity score in
  observational studies for causal effects}.
\newblock \bibinfo{journal}{\emph{Biometrika}} (\bibinfo{year}{1983}),
  \bibinfo{pages}{41--55}.
\newblock


\bibitem[\protect\citeauthoryear{Schnabel, Swaminathan, Singh, Chandak, and
  Joachims}{Schnabel et~al\mbox{.}}{2016}]%
        {schnabel2016recommendations}
\bibfield{author}{\bibinfo{person}{Tobias Schnabel}, \bibinfo{person}{Adith
  Swaminathan}, \bibinfo{person}{Ashudeep Singh}, \bibinfo{person}{Navin
  Chandak}, {and} \bibinfo{person}{Thorsten Joachims}.}
  \bibinfo{year}{2016}\natexlab{}.
\newblock \showarticletitle{Recommendations as treatments: Debiasing learning
  and evaluation}. In \bibinfo{booktitle}{\emph{ICML}}.
  \bibinfo{pages}{1670--1679}.
\newblock


\bibitem[\protect\citeauthoryear{Shen, Tan, and Zhai}{Shen
  et~al\mbox{.}}{2005}]%
        {shen2005implicit}
\bibfield{author}{\bibinfo{person}{Xuehua Shen}, \bibinfo{person}{Bin Tan},
  {and} \bibinfo{person}{ChengXiang Zhai}.} \bibinfo{year}{2005}\natexlab{}.
\newblock \showarticletitle{Implicit user modeling for personalized search}. In
  \bibinfo{booktitle}{\emph{CIKM}}. \bibinfo{pages}{824–831}.
\newblock


\bibitem[\protect\citeauthoryear{Sun, Khenissi, Nasraoui, and Shafto}{Sun
  et~al\mbox{.}}{2019}]%
        {sun2019debiasing}
\bibfield{author}{\bibinfo{person}{Wenlong Sun}, \bibinfo{person}{Sami
  Khenissi}, \bibinfo{person}{Olfa Nasraoui}, {and} \bibinfo{person}{Patrick
  Shafto}.} \bibinfo{year}{2019}\natexlab{}.
\newblock \showarticletitle{Debiasing the human-recommender system feedback
  loop in collaborative filtering}. In \bibinfo{booktitle}{\emph{Companion of
  WWW}}. \bibinfo{pages}{645--651}.
\newblock


\bibitem[\protect\citeauthoryear{Sun and Zhang}{Sun and Zhang}{2018}]%
        {sun2018conversational}
\bibfield{author}{\bibinfo{person}{Yueming Sun} {and} \bibinfo{person}{Yi
  Zhang}.} \bibinfo{year}{2018}\natexlab{}.
\newblock \showarticletitle{Conversational recommender system}. In
  \bibinfo{booktitle}{\emph{SIGIR}}. \bibinfo{pages}{235--244}.
\newblock


\bibitem[\protect\citeauthoryear{Tang, Niu, Huang, Shi, and Zhang}{Tang
  et~al\mbox{.}}{2020}]%
        {tang2020unbiased}
\bibfield{author}{\bibinfo{person}{Kaihua Tang}, \bibinfo{person}{Yulei Niu},
  \bibinfo{person}{Jianqiang Huang}, \bibinfo{person}{Jiaxin Shi}, {and}
  \bibinfo{person}{Hanwang Zhang}.} \bibinfo{year}{2020}\natexlab{}.
\newblock \showarticletitle{Unbiased scene graph generation from biased
  training}. In \bibinfo{booktitle}{\emph{CVPR}}. \bibinfo{pages}{3716--3725}.
\newblock


\bibitem[\protect\citeauthoryear{Wang, Huang, Zhang, and Sun}{Wang
  et~al\mbox{.}}{2020}]%
        {wang2020visual}
\bibfield{author}{\bibinfo{person}{Tan Wang}, \bibinfo{person}{Jianqiang
  Huang}, \bibinfo{person}{Hanwang Zhang}, {and} \bibinfo{person}{Qianru Sun}.}
  \bibinfo{year}{2020}\natexlab{}.
\newblock \showarticletitle{Visual commonsense r-cnn}. In
  \bibinfo{booktitle}{\emph{CVPR}}. \bibinfo{pages}{10760--10770}.
\newblock


\bibitem[\protect\citeauthoryear{Wang, Feng, He, Zhang, and Chua}{Wang
  et~al\mbox{.}}{2021}]%
        {wang2020click}
\bibfield{author}{\bibinfo{person}{Wenjie Wang}, \bibinfo{person}{Fuli Feng},
  \bibinfo{person}{Xiangnan He}, \bibinfo{person}{Hanwang Zhang}, {and}
  \bibinfo{person}{Tat-Seng Chua}.} \bibinfo{year}{2021}\natexlab{}.
\newblock \showarticletitle{" Click" Is Not Equal to" Like": Counterfactual
  Recommendation for Mitigating Clickbait Issue}. In
  \bibinfo{booktitle}{\emph{SIGIR}}.
\newblock


\bibitem[\protect\citeauthoryear{Wang, He, Wang, Feng, and Chua}{Wang
  et~al\mbox{.}}{2019}]%
        {wang2019ngcf}
\bibfield{author}{\bibinfo{person}{Xiang Wang}, \bibinfo{person}{Xiangnan He},
  \bibinfo{person}{Meng Wang}, \bibinfo{person}{Fuli Feng}, {and}
  \bibinfo{person}{Tat-Seng Chua}.} \bibinfo{year}{2019}\natexlab{}.
\newblock \showarticletitle{Neural Graph Collaborative Filtering}. In
  \bibinfo{booktitle}{\emph{SIGIR}}. \bibinfo{pages}{165--174}.
\newblock


\bibitem[\protect\citeauthoryear{Wang, Liang, Charlin, and Blei}{Wang
  et~al\mbox{.}}{2018}]%
        {wang2018deconfounded}
\bibfield{author}{\bibinfo{person}{Yixin Wang}, \bibinfo{person}{Dawen Liang},
  \bibinfo{person}{Laurent Charlin}, {and} \bibinfo{person}{David~M Blei}.}
  \bibinfo{year}{2018}\natexlab{}.
\newblock \showarticletitle{The deconfounded recommender: A causal inference
  approach to recommendation}.
\newblock \bibinfo{journal}{\emph{arXiv preprint arXiv:1808.06581}}
  (\bibinfo{year}{2018}).
\newblock


\bibitem[\protect\citeauthoryear{Wei, Wu, Li, Hu, Feng, He, Sun, and Wang}{Wei
  et~al\mbox{.}}{2020}]%
        {weifast}
\bibfield{author}{\bibinfo{person}{Tianxin Wei}, \bibinfo{person}{Ziwei Wu},
  \bibinfo{person}{Ruirui Li}, \bibinfo{person}{Ziniu Hu},
  \bibinfo{person}{Fuli Feng}, \bibinfo{person}{Xiangnan He},
  \bibinfo{person}{Yizhou Sun}, {and} \bibinfo{person}{Wei Wang}.}
  \bibinfo{year}{2020}\natexlab{}.
\newblock \showarticletitle{Fast Adaptation for Cold-Start Collaborative
  Filtering with Meta-Learning}. In \bibinfo{booktitle}{\emph{ICDM}}.
  \bibinfo{pages}{661--670}.
\newblock


\bibitem[\protect\citeauthoryear{Wu, Sun, Fu, Hong, Wang, and Wang}{Wu
  et~al\mbox{.}}{2019}]%
        {wu2019neural}
\bibfield{author}{\bibinfo{person}{Le Wu}, \bibinfo{person}{Peijie Sun},
  \bibinfo{person}{Yanjie Fu}, \bibinfo{person}{Richang Hong},
  \bibinfo{person}{Xiting Wang}, {and} \bibinfo{person}{Meng Wang}.}
  \bibinfo{year}{2019}\natexlab{}.
\newblock \showarticletitle{A neural influence diffusion model for social
  recommendation}. In \bibinfo{booktitle}{\emph{SIGIR}}.
  \bibinfo{pages}{235--244}.
\newblock


\bibitem[\protect\citeauthoryear{Xu, He, and Li}{Xu et~al\mbox{.}}{2020}]%
        {DL4match}
\bibfield{author}{\bibinfo{person}{Jun Xu}, \bibinfo{person}{Xiangnan He},
  {and} \bibinfo{person}{Hang Li}.} \bibinfo{year}{2020}\natexlab{}.
\newblock \showarticletitle{Deep Learning for Matching in Search and
  Recommendation}.
\newblock \bibinfo{journal}{\emph{Found. Trends Inf. Ret.}}
  \bibinfo{volume}{14} (\bibinfo{year}{2020}), \bibinfo{pages}{102--288}.
\newblock


\bibitem[\protect\citeauthoryear{Xue, Dai, Zhang, Huang, and Chen}{Xue
  et~al\mbox{.}}{2017}]%
        {xue2017deep}
\bibfield{author}{\bibinfo{person}{Hong-Jian Xue}, \bibinfo{person}{Xinyu Dai},
  \bibinfo{person}{Jianbing Zhang}, \bibinfo{person}{Shujian Huang}, {and}
  \bibinfo{person}{Jiajun Chen}.} \bibinfo{year}{2017}\natexlab{}.
\newblock \showarticletitle{Deep Matrix Factorization Models for Recommender
  Systems}. In \bibinfo{booktitle}{\emph{IJCAI}}. \bibinfo{pages}{3203--3209}.
\newblock


\bibitem[\protect\citeauthoryear{Ying, He, Chen, Eksombatchai, Hamilton, and
  Leskovec}{Ying et~al\mbox{.}}{2018}]%
        {ying2018graph}
\bibfield{author}{\bibinfo{person}{Rex Ying}, \bibinfo{person}{Ruining He},
  \bibinfo{person}{Kaifeng Chen}, \bibinfo{person}{Pong Eksombatchai},
  \bibinfo{person}{William~L Hamilton}, {and} \bibinfo{person}{Jure Leskovec}.}
  \bibinfo{year}{2018}\natexlab{}.
\newblock \showarticletitle{Graph convolutional neural networks for web-scale
  recommender systems}. In \bibinfo{booktitle}{\emph{KDD}}.
  \bibinfo{pages}{974--983}.
\newblock


\bibitem[\protect\citeauthoryear{Zhang, Jiang, Wang, Kuang, Zhao, Zhu, Yu,
  Yang, and Wu}{Zhang et~al\mbox{.}}{2020}]%
        {zhang2020devlbert}
\bibfield{author}{\bibinfo{person}{Shengyu Zhang}, \bibinfo{person}{Tan Jiang},
  \bibinfo{person}{Tan Wang}, \bibinfo{person}{Kun Kuang},
  \bibinfo{person}{Zhou Zhao}, \bibinfo{person}{Jianke Zhu},
  \bibinfo{person}{Jin Yu}, \bibinfo{person}{Hongxia Yang}, {and}
  \bibinfo{person}{Fei Wu}.} \bibinfo{year}{2020}\natexlab{}.
\newblock \showarticletitle{DeVLBert: Learning Deconfounded Visio-Linguistic
  Representations}. In \bibinfo{booktitle}{\emph{ACMMM}}.
  \bibinfo{pages}{4373--4382}.
\newblock


\bibitem[\protect\citeauthoryear{Zhang, Feng, He, Wei, Song, Ling, and
  Zhang}{Zhang et~al\mbox{.}}{2021}]%
        {zhang2021causal}
\bibfield{author}{\bibinfo{person}{Yang Zhang}, \bibinfo{person}{Fuli Feng},
  \bibinfo{person}{Xiangnan He}, \bibinfo{person}{Tianxin Wei},
  \bibinfo{person}{Chonggang Song}, \bibinfo{person}{Guohui Ling}, {and}
  \bibinfo{person}{Yongdong Zhang}.} \bibinfo{year}{2021}\natexlab{}.
\newblock \showarticletitle{Causal Intervention for Leveraging Popularity Bias
  in Recommendation}. In \bibinfo{booktitle}{\emph{SIGIR}}.
\newblock


\bibitem[\protect\citeauthoryear{Zheng, Gao, Li, He, Jin, and Li}{Zheng
  et~al\mbox{.}}{2021}]%
        {zheng2020dice}
\bibfield{author}{\bibinfo{person}{Yu Zheng}, \bibinfo{person}{Chen Gao},
  \bibinfo{person}{Xiang Li}, \bibinfo{person}{Xiangnan He},
  \bibinfo{person}{Depeng Jin}, {and} \bibinfo{person}{Yong Li}.}
  \bibinfo{year}{2021}\natexlab{}.
\newblock \showarticletitle{Disentangling User Interest and Conformity for
  Recommendation with Causal Embedding}. In \bibinfo{booktitle}{\emph{WWW}}.
\newblock


\bibitem[\protect\citeauthoryear{Zheng, Tang, Ding, and Zhou}{Zheng
  et~al\mbox{.}}{2016}]%
        {zheng2016neural}
\bibfield{author}{\bibinfo{person}{Yin Zheng}, \bibinfo{person}{Bangsheng
  Tang}, \bibinfo{person}{Wenkui Ding}, {and} \bibinfo{person}{Hanning Zhou}.}
  \bibinfo{year}{2016}\natexlab{}.
\newblock \showarticletitle{A Neural Autoregressive Approach to Collaborative
  Filtering}. In \bibinfo{booktitle}{\emph{ICML}}. \bibinfo{pages}{764–773}.
\newblock


\bibitem[\protect\citeauthoryear{Zhou, Zhu, Song, Fan, Zhu, Ma, Yan, Jin, Li,
  and Gai}{Zhou et~al\mbox{.}}{2018}]%
        {zhou2018deep}
\bibfield{author}{\bibinfo{person}{Guorui Zhou}, \bibinfo{person}{Xiaoqiang
  Zhu}, \bibinfo{person}{Chenru Song}, \bibinfo{person}{Ying Fan},
  \bibinfo{person}{Han Zhu}, \bibinfo{person}{Xiao Ma},
  \bibinfo{person}{Yanghui Yan}, \bibinfo{person}{Junqi Jin},
  \bibinfo{person}{Han Li}, {and} \bibinfo{person}{Kun Gai}.}
  \bibinfo{year}{2018}\natexlab{}.
\newblock \showarticletitle{Deep interest network for click-through rate
  prediction}. In \bibinfo{booktitle}{\emph{KDD}}. \bibinfo{pages}{1059--1068}.
\newblock


\end{thebibliography}
